\documentclass[prl,twocolumn,showpacs,nofootinbib,preprintnumbers,amsmath,amssymb,aps,longbibliography,superscriptaddress,floatfix]{revtex4-2}
\usepackage{graphicx,subfigure,epsfig}
\usepackage{dcolumn}
\usepackage{amssymb}
\usepackage{times}
\usepackage{amsmath}
\usepackage{amsfonts}
\usepackage{mathrsfs}
\usepackage{setspace}
\usepackage{latexsym}
\usepackage{bbm}
\usepackage{float}
\usepackage{flafter}
\usepackage{bm}
\usepackage{epstopdf}
\usepackage{color}
\usepackage{multirow}
\usepackage{physics} %Dirac bracket
\usepackage[export]{adjustbox}
\DeclareMathOperator{\sgn}{sgn}
\usepackage{braket}

\usepackage{tikz}                  % for the drawing

\usepackage{dsfont}

\usepackage{soul} % for underline with automatic line-break

\def \be {\begin{eqnarray}}
\def \ee {\end{eqnarray}}

\usepackage{xcolor}

\newcommand{\bea}{\begin{equation} \begin{aligned}}
\newcommand{\eea}{\end{aligned} \end{equation} }

\newcommand{\bpm}{\begin{pmatrix}}
\newcommand{\epm}{\end{pmatrix}}
\newcommand{\eps}{\epsilon}

\usepackage{xr}
\usepackage[pdftex,breaklinks,bookmarks=false,colorlinks,linkcolor=blue,citecolor=blue,urlcolor=blue,hypertexnames=false]{hyperref}

\usepackage[normalem]{ulem}

\usepackage{cleveref}
\crefname{appendix}{App.}{Apps.}
\crefname{equation}{Eq.}{Eqs.}
\crefname{figure}{Fig.}{Figs.}
\crefname{table}{Tab.}{Tabs.}
\crefname{section}{Sec.}{Secs.}
\creflabelformat{appendix}{#2#1#3}

\usepackage{orcidlink}

\begin{document}

\preprint{APS/123-QED}

\title{\textbf{1D Luttinger Modes in Carbon Nanotubes as keV Dark Matter Detector} 
}% 

\author{Xiuyuan Zhang}
\thanks{These two authors contributed equally.}
 \affiliation{Physics Department and Kavli Institute for Astrophysics and Space Research, \\Massachusetts Institute of Technology, Cambridge, MA 02139, USA}%Lines break automatically or can be forced with \\
  \email{Contact author: xiuyuan@mit.edu}
\author{Hao Chen}%
\thanks{These two authors contributed equally.}
\affiliation{Department of Electrical and Computer Engineering, Princeton University, Princeton, NJ 08540
}%
\affiliation{Department of Physics, Princeton University, Princeton, NJ 08540}

\date{\today}% It is always \today, today,

\begin{abstract}
We propose metallic carbon nanotubes (CNTs) as a one-dimensional plasmon target for light dark matter (DM) direct detection. 
Unlike conventional gapless electronic targets, where DM primarily excites electron-hole pairs, the low-energy charge response of a metallic CNT is carried by a collective Luttinger-liquid mode. 
We compute the projected sensitivity for DM-electron scattering through heavy and light mediators, using benchmark thresholds motivated by quantum-capacitance-detector-like and superconducting-quasiparticle-amplifying-transmon-like readout. 
For an accumulated nanotube length $L_{\text{CNT}}=10^8{\rm m}$, corresponding to milligram-scale single-wall CNT targets, we find competitive reach in the keV--MeV mass range. 
In the light-mediator case, the projected sensitivity can probe the cosmologically motivated freeze-in benchmark at keV masses. 
We also show that the one-dimensional geometry of aligned CNTs induces sidereal-day modulation, providing a handle for distinguishing a DM signal from approximately time-independent sensor backgrounds. 
These results establish one-dimensional collective modes as a new target class for sub-MeV DM detection. Existing progress in scalable CNT synthesis and superconducting quasiparticle sensing provides a promising experimental foundation, while realizing the proposed detector will require dedicated development of CNT--superconductor coupling and plasmon-to-quasiparticle conversion.

\end{abstract}

                              %display desired\maketitle
\maketitle
%\tableofcontents

The particle nature of DM remains one of the central open questions in particle physics and cosmology ~\cite{andernach2017englishspanishtranslationzwickys, 1980ApJ...238..471R, Bertone_2018}. Decades of direct-detection searches have placed stringent constraints on GeV-scale weakly interacting massive particle candidates ~\cite{PhysRevD.30.2295, PhysRevD.31.3059, PhysRevD.33.3495, Bo_2025, Aalbers_2025, Aprile_2025}, yet large regions of theoretically motivated parameter space remain unexplored, particularly at lower masses. In recent years, advances in low-threshold detector technologies and in the understanding of quantum materials have motivated a broad program to extend direct detection to the MeV and even keV mass range ~\cite{essig2016directdetectionsubgevdark, Hochberg_2016, Knapen_2018, PhysRevLett.123.151802, Baudis_2025}. 

Gapless electronic materials are especially promising in this context, since their low-energy excitations can be sensitive to the small energy deposits expected from sub-MeV DM~\cite{Hochberg_2018, Geilhufe_2020, das2024submevdarkmatterdetection, sherpa2026divedeepersubmarinesubmev}, potentially allowing experiments to approach the lowest masses consistent with phase-space constraints such as the Tremaine–Gunn bound~\cite{PhysRevLett.42.407}. Existing proposals have explored both three-dimensional Dirac materials~\cite{Hochberg_2018, Geilhufe_2020} and two-dimensional graphene-based systems~\cite{das2024submevdarkmatterdetection, sherpa2026divedeepersubmarinesubmev}, where the relevant electronic response is typically described in terms of electron-hole excitations near a linear band crossing or a small tunable gap. A one-dimensional conductor provides a qualitatively different gapless system. In one dimension, electron-electron interactions are strongly enhanced, and the low-energy theory is no longer described by independent electron-like quasiparticles. Instead, the appropriate description is a Luttinger liquid, whose elementary excitations are collective charge and spin density waves~\cite{33eb53d663e34ecc82e4e64ea6ca5cbe, Voit_1995}. The charge sector is characterized by a mode velocity $v$ and an interaction parameter $g$, with $g=1$ corresponding to the noninteracting limit and $g<1(g>1)$ describing repulsive (attractive) interactions. Since DM couples to electronic density, it can deposit energy and momentum by exciting these collective one-dimensional modes. Metallic CNTs naturally manifest this theoretical framework~\cite{PhysRevLett.79.5086, Egger_1997, Egger_1998}: they inherit gapless electronic modes from graphene, but the reduced dimensionality and long-range Coulomb interactions reorganize the low-energy response into Luttinger liquid charge modes rather than ordinary electron-hole excitations.

In this Letter, we propose the first ever experiment probing keV mass DM using 1D Luttinger mode in CNT which offers several key advantages. The collective response of CNTs allows the DM signal to be deposited into a coherent charge-density excitation, concentrating dynamical structure function into a well-defined propagating mode, $S(q_{\parallel}, \omega)\propto g |q_{\parallel}|\delta (\omega - v|q_{\parallel}|)$, rather than spreading it over a continuum of particle-hole states~\cite{PhysRevLett.79.5086, Egger_1997, Egger_1998, egger2000luttingerliquidbehaviormetallic, Ishii_2004, Imambekov_2012, Shi_2015}. The intrinsic anisotropy of a one-dimensional target can enhance directional dependence and produce daily modulation signals, providing an additional handle for separating a DM signal from isotropic backgrounds~\cite{PhysRevD.103.016006, PhysRevD.108.015015, das2024submevdarkmatterdetection}. In addition, CNTs also offers an important practical advantage: CNT synthesis and processing have been scaled to large-volume commercial production~\cite{Zhang_2011, Walker_2019, hughes_2024, NanoIntegris_Catalog_2020, NanoIntegris_MetallicSWNTs, SigmaAldrich_IsoNanotubesM_750530, Raymor_IsoNanotubesM, Graphenea_GrapheneProperties}. At the same time, the potential readout technologies are either already mature or under active development~\cite{2018NatAs...2...90E, Fink_2024}. This combination makes aligned metallic CNTs a particularly attractive near-term target: the material platform is substantially more developed than most proposed exotic quantum materials, while the readout can build on existing low-threshold superconducting detector technology. As a result, a CNT Luttinger-liquid detector may therefore offer a compelling route toward testing cosmologically motivated light-mediator benchmarks, including the freeze-in target for sub-MeV DM~\cite{PhysRevLett.127.111301}.

\textit{1D Luttinger Liquid Theory}--- Luttinger liquid theory describes low energy excitations in a 1D electron systems, for which CNT is a typical example. As a simplest model, the charge sector of a CNT is composed of one right-moving electron mode with annihilation operator $c_R(x)$ and one left-moving electron mode with annihilation operator $c_L(x)$, each having a linear dispersion relation. We consider the following Hamiltonian
\begin{equation}
    H = \int dx\mathcal{H}(x),\ 
    \mathcal{H}(x) = -iv_F c_R^\dag\partial_xc_R + iv_Fc_L^\dag\partial_xc_L + \mathcal{H}_{\text{int}},
\end{equation}
where the two modes have velocities $\pm v_F$ with $v_F > 0$, respectively. In the interaction part $\mathcal{H}_{\text{int}}$ of the Hamiltonian density, we include short-range density-density interactions between electrons, given by the forward and backward scattering terms:
\begin{equation}
    \mathcal{H}_{\text{int}}(x) = \frac{g_4}{2}\left[n_R^2(x)+n_L^2(x)\right] + g_2n_R(x)n_L(x)
\end{equation}
where the density operators $n_{R,L}(x) = c_{R,L}^\dag(x)c_{R,L}(x)$, and the real parameter $g_4$ ($g_2$) is the strength of the forward (backward) scattering. This model, though quartic in fermions, is exactly solvable via bosonization \cite{bosonization,bosonization1,fradkin2013field} due to its 1D character, leading to a closed-form expression of the density-density correlation function (see End Matter)
\begin{equation}\label{eq: nn_corr}
    \braket{n(x,t) n(0,0)} = -\frac{g}{4\pi^2}\left[\frac{1}{(x-vt+i\epsilon)^2}+\frac{1}{(x+vt-i\epsilon)^2}\right],
\end{equation}
where we have defined the Luttinger parameter $g = \sqrt{\frac{2\pi v_F+g_4-g_2}{2\pi v_F+g_4+g_2}}$ and the velocity $v=\sqrt{\left(v_F+\frac{g_4}{2\pi}\right)^2-\left(\frac{g_2}{2\pi}\right)^2}$, and $g<1$ ($g>1$) indicates the repulsive (attractive) interaction between left and right moving modes.
The corresponding structure factor in 1D is then
\begin{equation}
\begin{aligned}
    S_{\text{1D}}(q,\omega) &= \iint dtdx\  e^{i(\omega t-qx)}\braket{n(x,t) n(0,0)}\\ &=  g |q| \delta(\omega - v|q|).
\end{aligned}
\end{equation}

What is needed for evaluating the scattering rate between DM and CNT is the 3D structure factor $S_{\text{3D}}(\mathbf{q},\omega) = \int dt e^{i\omega t}\braket{n_{\text{3D}}(\mathbf{q},t) n_{\text{3D}}(-\mathbf{q},0)}$, where $n_{\text{3D}}(\mathbf{q},t)$ is the Fourier transform of 3D electron density $n_{\text{3D}}(\textbf{r}) = n_{\text{3D}}(r_\parallel,r_\perp,\theta) = n_{\text{1D}}(r_\parallel)f_\perp(r_\perp,\theta)$ at time $t$. Here, we use $r_\parallel$ use the coordinate along the CNT and use $\mathbf{r}_\perp = (r_\perp,\theta)$ to denote the radial and angular coordinates in perpendicular to the CNT, while we take $f_\perp(r_\perp,\theta) = \frac{\delta(r_\perp-r_0)}{2\pi r_0}$ as the charge density profile in perpendicular to the CNT of radius $r_0$. In this way, $n_{\text{3D}}(\mathbf{q},t) = n_{\text{1D}}(q_\parallel,t)F_\perp(\mathbf{q}_\perp)$, and  $F_\perp(\mathbf{q}_\perp) = \int d^2r_\perp e^{-i\mathbf{q}_\perp \cdot \mathbf{r}_\perp} f_\perp(\mathbf{r}_\perp) = J_0(q_\perp r_0)$ bears the shape of the zeroth-order Bessel function $J_0$, with $q_\perp = |\mathbf{q}_\perp|$~\cite{PhysRevB.64.241402, Hamley_2023}.

\textit{DM scattering rate}---We consider DM--electron scattering mediated by a Yukawa interaction. For a density-coupled interaction, the scattering rate is controlled by the dynamical structure factor of the target. In the CNT system considered here, the relevant low-energy density response is from the one-dimensional Luttinger-liquid charge modes derived in the previous section. We therefore use this response to estimate the rate for DM to excite CNT Luttinger modes, providing a proof-of-principle calculation for their use as light DM targets. The general scattering rate can be written as

\begin{equation}
\begin{split}
R&=\frac{\rho_{\chi}}{m_{\chi}}\int f(\mathbf{v}_\chi) d\mathbf{v}_\chi\int\frac{d^3q}{(2\pi)^3} |V(\mathbf{q})|^2S_{\text{3D}}(\mathbf{q}, \omega)\\
&=\frac{\rho_{\chi}}{m_{\chi}}\frac{\pi \bar{\sigma}_{\chi e}}{\mu_{\chi e}^2}\int f(\mathbf{v}_\chi) d\mathbf{v}_\chi\int\frac{d^3q}{(2\pi)^3} F_{\chi}(\mathbf{q})^2S_{\text{3D}}(\mathbf{q}, \omega)
\end{split}
\end{equation}
where $m_{\chi}$ and $\rho_{\chi}$ are mass and local density of the DM species, and $f(\mathbf{v}_\chi)$ is the velocity distribution of DM. $|V(\mathbf{q})|^2=g_e^2g_{\chi}^2/(q^2+m_{\phi}^2)^2=\pi\bar{\sigma}_{\chi e}F_{\chi}^2(\mathbf{q})/\mu_{\chi e}^2$ is the non-relativistic potential with a mediator mass $m_{\phi}$ and cross section $\bar{\sigma}_{\chi e}$. $F_{\chi}(\mathbf{q}) = (\alpha^2 m_e^2 + m_{\phi}^2)/(q^2 + m_{\phi}^2)$ is a form factor characterizing the momentum transfer dependence. The 3D dynamical structure $S_{\text{3D}}(\mathbf{q}, \omega) = \frac{1}{V}\int dt e^{i \omega t} \langle \hat{n}_{\text{3D}}(\mathbf{q}, t), \hat{n}_{\text{3D}}(\mathbf{-q}, 0) \rangle$ is related to $S_{\text{1D}}(q_{\parallel},\omega)$ through a projection on the density operators: $\hat{n}_{\text{3D}}(\mathbf{q}) = F_{\perp}(\mathbf{q}_{\perp})\hat{n}_{\text{1D}}(q_{\parallel})$. Thus the scattering rate can be written as~\cite{essig2016directdetectionsubgevdark, Trickle_2020, Hochberg_2021, Knapen_2021}: 

\begin{equation}
\begin{split}
\frac{R}{L_{\text{CNT}}}=\frac{\rho_{\chi}}{m_{\chi}}&\frac{\pi \bar{\sigma}_{\chi e}}{\mu_{\chi e}^2}\int\frac{d^3q}{(2\pi)^3} F_{\chi}(\mathbf{q})^2|F_{\perp}(\mathbf{q}_{\perp})|^2g  |\mathbf{q}_{\parallel}| \Theta(\mathbf{q}_{\parallel})\\
&\times \int f(\mathbf{v}_\chi) d\mathbf{v}_\chi\delta(\mathbf{q} \cdot \mathbf{v}_\chi -\frac{q^2}{2m_{\chi}}- v \mathbf{q}_{\parallel}))
\end{split}
\end{equation}

The one-dimensional nature of the target also affects how the rate should be normalized. In conventional bulk targets, event rates are usually quoted in units of counts per unit time per unit detector mass, reflecting the scaling of the active target with volume. In contrast, the relevant active degree of freedom in the present system is the accumulated length of one-dimensional CNT modes. We therefore report the scattering rate per unit nanotube length, $R/L_{\rm CNT}$.

For the DM velocity distribution $f(\mathbf{v}_\chi)$, we assume the Standard Halo Model, with a Maxwellian distribution truncated at the Galactic escape velocity and boosted into the laboratory frame~\cite{PhysRevD.30.2295}. We take $v_0=220{\rm km/s}, v_{\rm esc}=550{\rm km/s}$~\cite{2024ApJ...972...70R}, and a characteristic lab-frame velocity $v_E=232{\rm km/s}$~\cite{PhysRevD.33.3495, PhysRevD.37.3388}. 
%Because the CNT response is intrinsically anisotropic, the orientation of the nanotube axis relative to the incoming DM wind changes as the Earth rotates. This induces a sidereal-day modulation of the signal rate, providing a directional handle that is absent or less pronounced in isotropic targets.

\begin{figure}[t]
    \centering
    \includegraphics[width=1\linewidth]{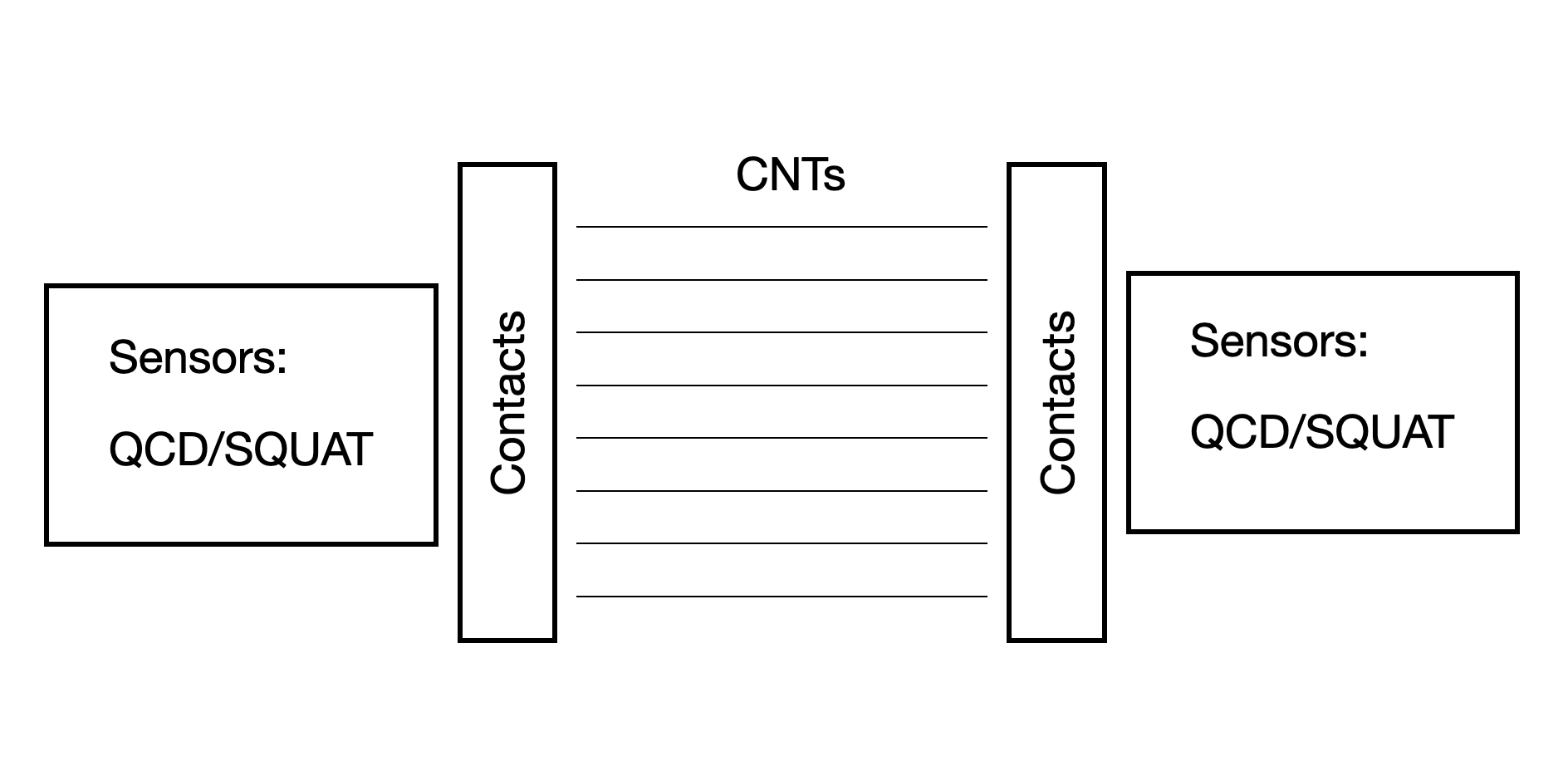}
    \caption{Schematic of the proposed CNT detector architecture. Each pixel contains an aligned forest of metallic SWCNTs connected in parallel between superconducting absorber contacts integrated with quasiparticle-sensitive readout sensors.
    }
    \label{fig:design}
\end{figure}

\begin{figure*}[t]
    \centering
    \includegraphics[width=0.45\linewidth]{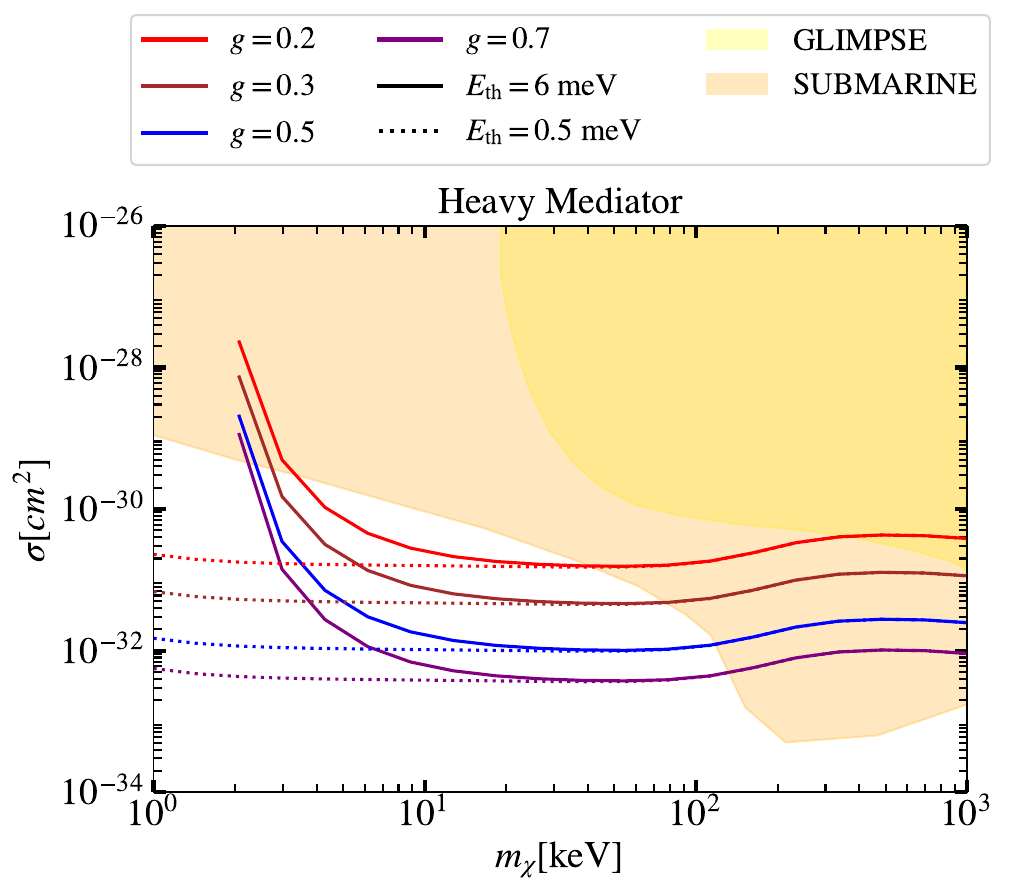}
    \includegraphics[width=0.45\linewidth]{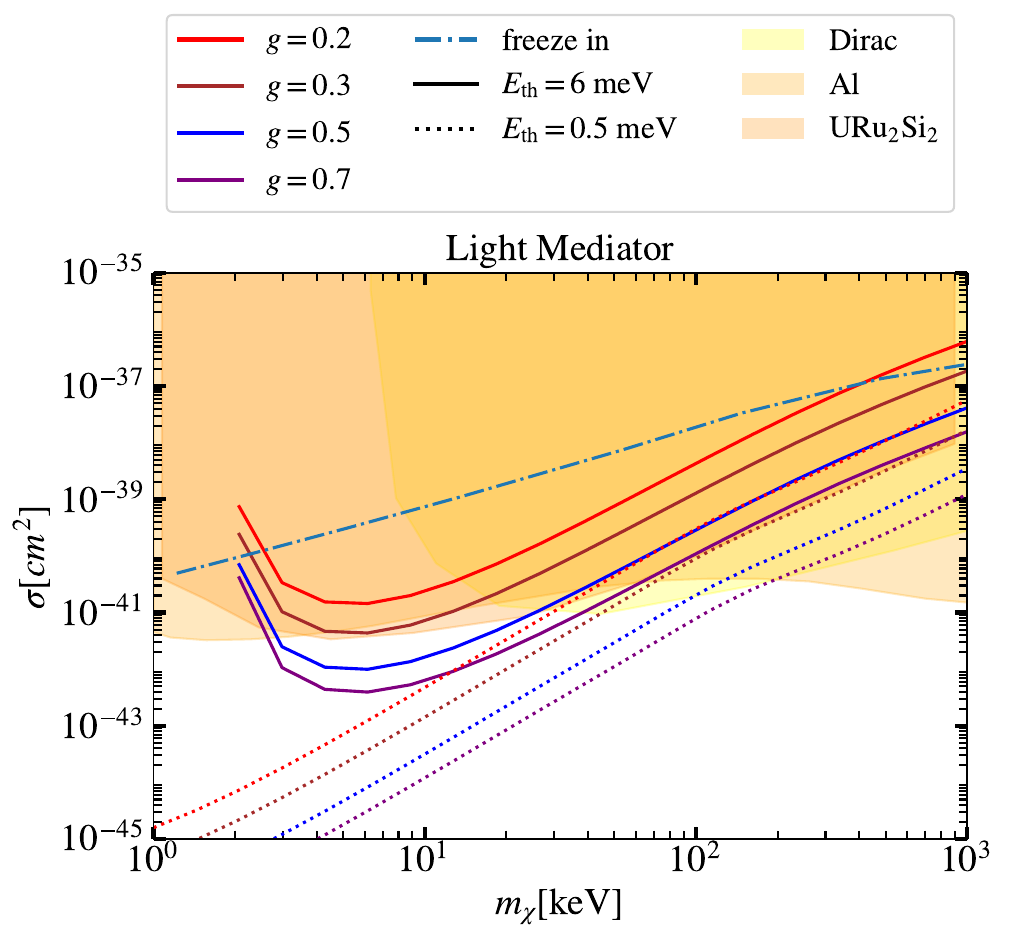}
    \caption{
    Projected sensitivity of a CNT Luttinger-liquid DM detector to DM--electron scattering. The left panel shows the heavy-mediator benchmark ($F_\chi(q)=1$), while the right panel shows the light-mediator benchmark ($F_\chi(q)=\alpha^2m_e^2/q^2$). Solid curves correspond to a readout threshold $E_{\rm th}=6{\rm meV}$, and dotted curves correspond to an optimistic threshold $E_{\rm th}=0.5{\rm meV}$. In the heavy-mediator case, we compare to graphene-based proposals including GLIMPSE and SUBMARINE. In the light-mediator case, we compare to far-term material benchmarks from Ref.~\cite{Hochberg_2021}, including aluminum, a benchmark Dirac material, and $\rm URu_2Si_2$, together with the freeze-in target shown by the dash-dotted curve. 
    }
    \label{fig:projection}
\end{figure*}

\textit{Detectors and projected sensitivity}---We are ready to discuss the detector design and compute the projected sensitivity. Our proposed design is schematically shown in Fig.~\ref{fig:design}. Each detector pixel consists of a forest of metallic single-walled carbon nanotubes (SWCNTs) connected in parallel between two absorber contacts, which are subsequently coupled to the readout sensors. In terms of total amount of the target material, mg-scale metallic SWCNTs has been made accessible, with g/kg quotes available by request at several vendors~\cite{NanoIntegris_Catalog_2020, NanoIntegris_MetallicSWNTs, SigmaAldrich_IsoNanotubesM_750530, Raymor_IsoNanotubesM, Graphenea_GrapheneProperties}. Since the event rate only cares about the total length, we need to convert the mass scale to length scale. For commercial SWCNTs, the diameter is usually $d\simeq 1.2-1.7$nm. Thus the linear mass density is roughly $\lambda \simeq \rho_{\rm graphene}\pi d\simeq (2.9-4.1)\times 10^{-12}$g/m. So a mg-scale sample will translate into an accumulated length of $10^8$m, which we will choose as a benchmark. 

Another important scale is the energy threshold determined by the readout scheme. Here we consider two different potential readout schemes as near-term and far-term benchmarks, quantum-capacitance-detector-like (QCD-like) and superconducting-quasiparticle-amplifying-transmon-like (SQUAT-like). A QCD is a superconducting quasiparticle sensor in which a small energy deposition creates quasiparticles that tunnel onto a superconducting island, changing its quantum capacitance and shifting the response of a microwave resonator~\cite{2018NatAs...2...90E}. QCDs have so far been developed primarily for far-infrared and THz photon detection with an energy threshold of $E_{\text{th}}=6$meV, which we choose as our near-term benchmark due to the maturity of the technology.  

SQUAT is a proposed qubit-based sensor in which a small energy deposition creates quasiparticles in a superconducting transmon circuit; quasiparticle tunneling changes the qubit parity, producing a measurable charge-sensitive signal~\cite{Fink_2024, magoon2026demonstrationsquatdetectorarchitecture}. The architecture was designed for ultra-low-threshold detection of meV-scale phonons and single THz photons. Unlike QCD, SQUAT is still at an early development stage. A 2026 preprint~\cite{magoon2026demonstrationsquatdetectorarchitecture} reports a first demonstration of the SQUAT detector architecture with resonator-free charge-sensitive transmons and simultaneous charge/quasiparticle signals in aluminum devices. Thus we show an optimistic SQUAT-like benchmark at $E_{\text{th}}=0.5$meV, motivated by the sub-meV pair-breaking scale of low-gap superconductors~\cite{ramanathan2026quantumparitydetectorsqubit} and the rapid development of quasiparticle-amplifying transmon sensors as a far-term benchmark. 

It should be noticed that neither QCD nor SQUAT was designed to couple directly with plasmon modes. Clean CNT devices have already been coupled to superconducting microwave circuits~\cite{Ranjan_2015, PhysRevApplied.15.064050}, and Luttinger-liquid plasmons have been observed in metallic nanotubes using infrared and THz techniques~\cite{2013APS..MAR.N8001C, Shi_2015}. An additional coupling structure is therefore required to transfer the CNT plasmon energy into the superconducting absorber and convert it into quasiparticle excitations. The material composition and geometry of this interface must be engineered carefully. In particular, it should present an effective impedance comparable to the characteristic impedance of the CNT plasmon mode, which is expected to be of order $10\mathrm{k}\Omega$~\cite{Burke_2002_Luttinger, Burke_2003_RF}, in order to suppress reflection at the CNT termination. At the same time, the dominant dissipative channel within the coupling structure should be Cooper-pair breaking, rather than transmission into other electromagnetic modes, radiation, substrate excitations, or ordinary thermal losses. Achieving both efficient impedance matching and a large plasmon-to-quasiparticle conversion efficiency will require dedicated detector-development and characterization efforts.

The length of each CNT tube should not be much longer than the propagation length of the plasmon mode. Although the propagation length ultimately depends on the plasmon damping rate and must therefore be measured for the fabricated CNT target, the plasmon wavelength provides a useful geometric reference scale. Within the Luttinger-liquid description, the group velocity of the plasmon mode is $v_g=v_F/g$, where $g$ is a target-dependent quantity that should ultimately be measured for the fabricated CNT sample. Existing theoretical estimates and experimental measurements in metallic nanotubes suggest values in the approximate range $0.2\sim 0.7$~\cite{PhysRevLett.79.5086, Bockrath_1999, 2013APS..MAR.N8001C}. We therefore present benchmark sensitivities for $g=0.2, 0.3, 0.5, \text{and }  0.7$. Taking $v_F\simeq0.8\times10^6\mathrm{m/s}$, our benchmarks give $v_g\sim 2.5\times 10^6\mathrm{m/s}$. For an energy threshold $E_{\text{th}}=0.5\mathrm{meV}$, the corresponding frequency is $f=E_{\text{th}}/h\approx 121\mathrm{Hz} $ and hence $\lambda = v_g/f\sim 10\mu \mathrm{m}$. We therefore assume each CNT tube to be around $10\mu \mathrm{m}$ long, which corresponds to $N_{\rm CNT} \sim 10^{13}$ individual CNT tubes. 

A separate consideration is the spacing between neighboring nanotubes. If the nanotubes are packed too densely, intertube Coulomb interactions, screening, and collective plasmon modes may modify the response relative to that of an isolated nanotube. For nonrelativistic DM, the characteristic momentum transfer scales as $q\sim m_\chi v_\chi$, with $v_\chi/c\sim10^{-3}$ and $c$ being the speed of light. A representative $m_\chi\sim1\mathrm{keV}$ particle therefore transfers momentum of order $q\sim1\mathrm{eV}$, corresponding to an inverse momentum scale $\frac{\hbar c}{q}\sim200\mathrm{nm}$. If suppressing collective intertube effects requires transverse spacings of order $a\sim100\text{--}200\mathrm{nm}$, then $10^{13}$ nanotubes would occupy a total transverse area of approximately $A_{\rm det}\sim N_{\rm CNT}a^2\sim0.1\text{--}0.4\mathrm{m^2}$. Such a footprint would pose substantial challenges for superconducting-device fabrication. Because the fabrication capability is also strongly motivated by other quantum-sensing, detector, and industrial applications, substantial progress is expected independently of the present proposal. A smaller-scale first-generation detector may therefore be pursued while large-area superconducting fabrication continues to mature.

The projected sensitivity is calculated assuming a background free experiment at $95\%$ Confidence Level, which corresponds to 3 events per year. We will assume a local DM density of $0.4\rm{GeV/cm^3}$~\cite{Riccardo_Catena_2010, de_Salas_2021}. The results for heavy and light mediators are shown in Fig.~\ref{fig:projection}. Solid curves correspond to a benchmark energy threshold of $E_{\rm th}=6{\rm meV}$, motivated by QCD-like readout, while dotted curves correspond to an optimistic $E_{\rm th}=0.5{\rm meV}$ SQUAT-like benchmark. For the heavy-mediator case, shown in the left panel, we compare the projected reach of the CNT Luttinger-liquid target with existing graphene-based proposals, including GLIMPSE and SUBMARINE. Since GLIMPSE presents results only up to $\mu{\rm g}\cdot{\rm yr}$ exposures, we rescale their sensitivity to the corresponding ${\rm mg}\cdot{\rm yr}$ exposure for comparison, assuming background-free exposure scaling. GLIMPSE proposes a graphene Josephson-junction sensor with an energy threshold of $0.1{\rm meV}$, while SUBMARINE assumes a $50{\rm meV}$ threshold without specifying a detailed readout implementation. Relative to these graphene-based benchmarks, the CNT proposal achieves stronger projected sensitivity in the $m_\chi\sim10-100{\rm keV}$ mass range for the $6{\rm meV}$ threshold, and more broadly below $m_\chi\sim 100{\rm keV}$ for the optimistic $0.5{\rm meV}$ threshold. This improvement arises from the collective one-dimensional density response of the CNT Luttinger mode, together with the large accumulated nanotube length accessible in a scalable CNT material platform.

The light-mediator projections are shown in the right panel of Fig.~\ref{fig:projection}. We compare the CNT reach with the far-term material benchmarks studied in Ref.~\cite{Hochberg_2021}, including an aluminum superconducting target, a benchmark three-dimensional Dirac material, and $\rm URu_2Si_2$. These materials represent qualitatively different low-energy electronic responses. In aluminum, the projected sensitivity is controlled by the low-energy tail of the plasmon response in a conventional superconductor. In the Dirac-material benchmark, the reach is instead determined by interband excitations across a small gap, with the low Fermi velocity improving kinematic matching to sub-MeV DM. $\rm URu_2Si_2$ provides a complementary strongly correlated target whose measured loss function contains low-energy spectral features in the meV range. The CNT proposal differs from all of these cases in that its dynamical structure factor is concentrated in a propagating collective excitation.

For a light mediator, the scattering rate is strongly weighted toward small momentum transfer, making the low-$q$, low-$\omega$ density response of the target especially important. The CNT Luttinger mode therefore provides a distinct route to the same light-mediator parameter space probed by other benchmarks. We find that, for the assumed accumulated nanotube length and readout thresholds, the CNT detector can probe the freeze-in target in the keV-mass range and achieve sensitivity comparable to these far-term material proposals. This comparison suggests that aligned metallic CNTs could provide a competitive light-mediator target, with the additional practical advantage of a scalable one-dimensional material platform and readout thresholds motivated by superconducting quasiparticle sensors.

\begin{figure}[t]
    \centering
    \includegraphics[width=1\linewidth]{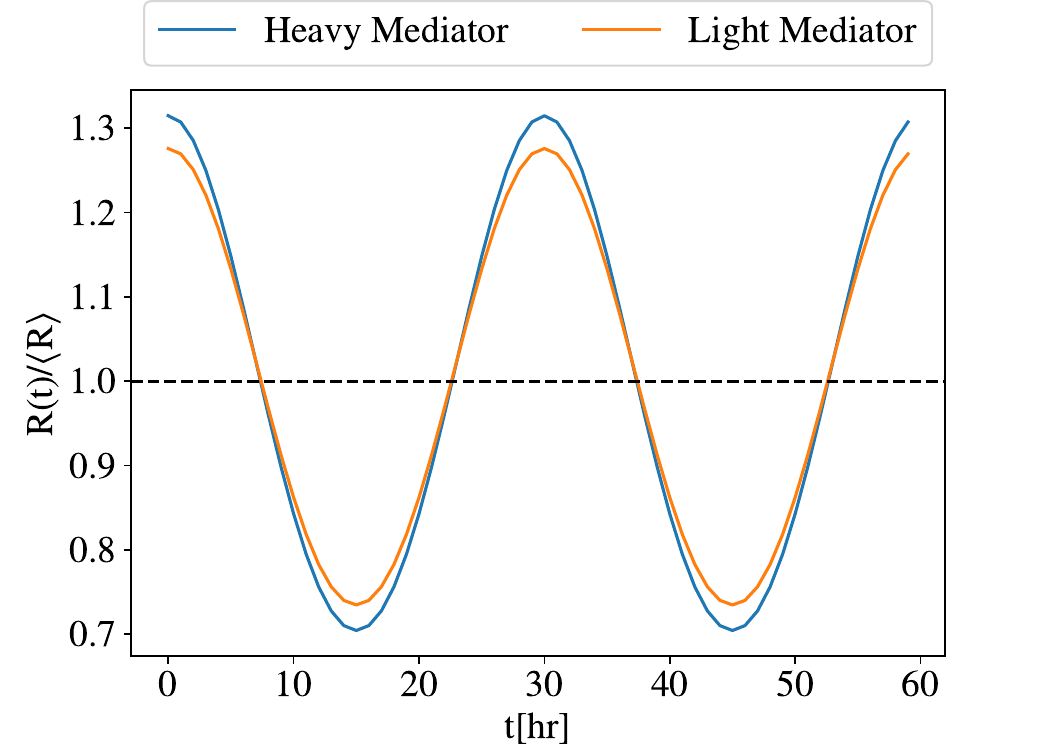}
    \caption{
    Daily modulation of the CNT Luttinger-liquid scattering rate for representative heavy- and light-mediator benchmarks. The rate is shown as a function of sidereal time for $m_\chi=10{\rm keV}$, $g=0.3$, and $E_{\rm th}=6{\rm meV}$. We define the modulation fraction as $A=(R_{\rm max}-R_{\rm min})/(R_{\rm max}+R_{\rm min})$, and find a characteristic amplitude $A\simeq 0.3$ for the benchmark parameters shown. 
    }
    \label{fig:DM}
\end{figure}

\textit{Backgrounds and daily modulation}---The typical backgrounds for a cryogenic CNT-based quasiparticle readout are expected to arise from thermal excitations, external particle interactions, and sensor dark triggers. Equilibrium thermal excitations are strongly Boltzmann suppressed at dilution-refrigerator temperatures: for $T\sim 50{\rm mK}$, $k_BT\simeq 4\times 10^{-3}{\rm meV}$, making meV-scale CNT plasmon excitations negligible~\cite{2018NatAs...2...90E}. External particle backgrounds from cosmic rays and environmental radioactivity can be reduced using standard shielding, vetoes, and underground operation~\cite{PhysRevD.73.053004, Li_2025, akerib2026measurementmuonfluxsanford}. The remaining instrumental background is expected to be dominated by nonequilibrium quasiparticles in the superconducting absorber, which produce spurious tunneling and parity-switching events. The QCD exhibited a residual quasiparticle tunneling-in rate of approximately $8\mathrm{kHz}$~\cite{2018NatAs...2...90E}, whereas preliminary measurements of later-generation SQUAT devices report quiescent parity-switching rates of order $3\text{--}10\mathrm{Hz}$~\cite{Aralis_2026_SQUAT}. These raw switching rates should not, however, be identified directly with event-level background rates. A signal energy deposition is expected to generate a transient enhancement of the quasiparticle population and hence a temporally localized burst of tunneling events~\cite{ramanathan2026quantumparitydetectorsqubit}. Requiring multiple switches within an appropriately chosen signal window can therefore strongly suppress accidental backgrounds from isolated quiescent transitions. In addition, the predicted daily modulation of the DM signal provides a further handle against time-independent or laboratory-correlated backgrounds.

To illustrate the modulation signal, in Fig.~\ref{fig:DM} we show the predicted daily modulation for representative heavy- and light-mediator benchmarks with $m_\chi=10{\rm keV}$, g=0.3, and $E_{\rm th}=6{\rm meV}$. For this benchmark, we choose the aligned CNT axis to be perpendicular to the Earth's rotation axis, which maximizes the change in the angle between the one-dimensional target and the DM wind over a sidereal day. We define the modulation fraction as $A \equiv \frac{R_{\rm max}-R_{\rm min}}{R_{\rm max}+R_{\rm min}}$, and find a characteristic amplitude $A\simeq 0.3$ for these benchmark parameters. If the total time-averaged signal and background counts over the exposure are denoted by S and B, respectively, the modulated component of the signal is of order AS, while the sensor and detector background contributes primarily to the unmodulated rate. A template analysis using the known sidereal phase can therefore search for the modulated component with an approximate significance $Z_{\rm mod}\sim \frac{A S}{\sqrt{2(S+B)}}$, or $Z_{\rm mod}\sim A S/\sqrt{2B}$ in the background-dominated limit. Up to order-one factors that depend on the time binning and modulation template, this shows that daily modulation can provide a powerful handle for distinguishing a CNT DM signal from unmodulated backgrounds. A detailed background model will ultimately be required in the experiment, but these considerations suggest that the leading reducible backgrounds will not overwhelm the projected signal.

\textit{Conclusion}---In this Letter, we have proposed metallic CNTs as a one-dimensional Luttinger-liquid target for light DM detection. The key feature of this system is that the low-energy electronic response is carried by a collective charge-density mode, which provides a distinct material response for density-coupled DM scattering, with strong intrinsic anisotropy and an associated sidereal-day modulation that can help distinguish a signal from approximately time-independent sensor backgrounds. Using benchmark readout thresholds motivated by demonstrated QCD-like and actively developing SQUAT-like superconducting sensors, we find that an accumulated nanotube length corresponding to milligram-scale SWCNT targets can achieve competitive sensitivity to keV-MeV DM. In particular, for a light mediator, the CNT detector can reach the cosmologically motivated freeze-in benchmark in the keV-mass range, with sensitivity comparable to other far-term quantum-material proposals. These results establish one-dimensional collective modes as a promising new target class for sub-MeV dark-matter detection. Existing technologies in CNT synthesis, purification, and alignment provide a useful experimental foundation, although realizing the proposed detector will require dedicated development of CNT--superconductor coupling, plasmon-to-quasiparticle conversion, target integration, and low-background readout.

\begin{acknowledgments}
We would like to thank Ben Lehmann, Gonzalo Herrera, Biao Lian, Lina Necib, Tracy Slayter, Pok Man Tam for helpful conversations.
X.Z. is supported by DOE grant DE-SC0024112.
H.C. is supported by the National Science Foundation under Award No. DMR-2141966, and receives additional supports from Bede Liu Fund for Excellence at the Department of Electrical and Computer Engineering of Princeton University. 
\end{acknowledgments}

\nocite{*}

\bibliography{apssamp}% Produces the bibliography via BibTeX.

@misc{bosonization,
Author = {D. Sénéchal},
Title = {An introduction to bosonization},
Year = {1999},
Eprint = {arXiv:cond-mat/9908262},
}

@article{bosonization1,
author = {von Delft, Jan and Schoeller, Herbert},
title = {Bosonization for beginners — refermionization for experts},
journal = {Annalen der Physik},
volume = {510},
number = {4},
pages = {225-305},
doi = {https://doi.org/10.1002/andp.19985100401},
url = {https://onlinelibrary.wiley.com/doi/abs/10.1002/andp.19985100401},
eprint = {https://onlinelibrary.wiley.com/doi/pdf/10.1002/andp.19985100401},
year = {1998}
}

@book{fradkin2013field,
  title={Field theories of condensed matter physics},
  author={Fradkin, Eduardo},
  year={2013},
  publisher={Cambridge University Press}
}

@misc{andernach2017englishspanishtranslationzwickys,
      title={English and Spanish Translation of Zwicky's (1933) The Redshift of Extragalactic Nebulae}, 
      author={Heinz Andernach and Fritz Zwicky},
      year={2017},
      eprint={1711.01693},
      archivePrefix={arXiv},
      primaryClass={astro-ph.IM},
      url={https://arxiv.org/abs/1711.01693}, 
}

@ARTICLE{1980ApJ...238..471R,
       author = {{Rubin}, V.~C. and {Ford}, Jr., W.~K. and {Thonnard}, N.},
        title = "{Rotational properties of 21 SC galaxies with a large range of luminosities and radii, from NGC 4605 (R=4kpc) to UGC 2885 (R=122kpc).}",
      journal = {\apj},
         year = 1980,
        month = jun,
       volume = {238},
        pages = {471-487},
          doi = {10.1086/158003},
       adsurl = {https://ui.adsabs.harvard.edu/abs/1980ApJ...238..471R}
}

@article{Bertone_2018,
   title={History of dark matter},
   volume={90},
   ISSN={1539-0756},
   url={http://dx.doi.org/10.1103/RevModPhys.90.045002},
   DOI={10.1103/revmodphys.90.045002},
   number={4},
   journal={Reviews of Modern Physics},
   publisher={American Physical Society (APS)},
   author={Bertone, Gianfranco and Hooper, Dan},
   year={2018},
   month=Oct }

@article{PhysRevD.31.3059,
  title = {Detectability of certain dark-matter candidates},
  author = {Goodman, Mark W. and Witten, Edward},
  journal = {Phys. Rev. D},
  volume = {31},
  issue = {12},
  pages = {3059--3063},
  numpages = {0},
  year = {1985},
  month = {Jun},
  publisher = {American Physical Society},
  doi = {10.1103/PhysRevD.31.3059},
  url = {https://link.aps.org/doi/10.1103/PhysRevD.31.3059}
}

@article{PhysRevD.30.2295,
  title = {Principles and applications of a neutral-current detector for neutrino physics and astronomy},
  author = {Drukier, A. and Stodolsky, L.},
  journal = {Phys. Rev. D},
  volume = {30},
  issue = {11},
  pages = {2295--2309},
  numpages = {0},
  year = {1984},
  month = {Dec},
  publisher = {American Physical Society},
  doi = {10.1103/PhysRevD.30.2295},
  url = {https://link.aps.org/doi/10.1103/PhysRevD.30.2295}
}

@article{PhysRevD.33.3495,
  title = {Detecting cold dark-matter candidates},
  author = {Drukier, Andrzej K. and Freese, Katherine and Spergel, David N.},
  journal = {Phys. Rev. D},
  volume = {33},
  issue = {12},
  pages = {3495--3508},
  numpages = {0},
  year = {1986},
  month = {Jun},
  publisher = {American Physical Society},
  doi = {10.1103/PhysRevD.33.3495},
  url = {https://link.aps.org/doi/10.1103/PhysRevD.33.3495}
}

@article{Aalbers_2025,
   title={Dark Matter Search Results from 4.2 Tonne-Years of Exposure of the {LUX-ZEPLIN} ({LZ}) Experiment},
   volume={135},
   ISSN={1079-7114},
   url={http://dx.doi.org/10.1103/4dyc-z8zf},
   DOI={10.1103/4dyc-z8zf},
   number={1},
   journal={Physical Review Letters},
   publisher={American Physical Society (APS)},
   author={Aalbers, J. and Akerib, D. S. and Musalhi, A. K. Al and Alder, F. and Amarasinghe, C. S. and Ames, A. and Anderson, T. J. and Angelides, N. and Araújo, H. M. and Armstrong, J. E. and Arthurs, M. and Baker, A. and Balashov, S. and Bang, J. and Bargemann, J. W. and Barillier, E. E. and Bauer, D. and Beattie, K. and Benson, T. and Bhatti, A. and Biekert, A. and Biesiadzinski, T. P. and Birch, H. J. and Bishop, E. and Blockinger, G. M. and Boxer, B. and Brew, C. A. J. and Brás, P. and Burdin, S. and Buuck, M. and Carmona-Benitez, M. C. and Carter, M. and Chawla, A. and Chen, H. and Cherwinka, J. J. and Chin, Y. T. and Chott, N. I. and Converse, M. V. and Coronel, R. and Cottle, A. and Cox, G. and Curran, D. and Dahl, C. E. and Darlington, I. and Dave, S. and David, A. and Delgaudio, J. and Dey, S. and de Viveiros, L. and Di Felice, L. and Ding, C. and Dobson, J. E. Y. and Druszkiewicz, E. and Dubey, S. and Eriksen, S. R. and Fan, A. and Fayer, S. and Fearon, N. M. and Fieldhouse, N. and Fiorucci, S. and Flaecher, H. and Fraser, E. D. and Fruth, T. M. A. and Gaitskell, R. J. and Geffre, A. and Genovesi, J. and Ghag, C. and Ghosh, A. and Gibbons, R. and Gokhale, S. and Green, J. and van der Grinten, M. G. D. and Haiston, J. J. and Hall, C. R. and Hall, T. J. and Han, S. and Hartigan-O’Connor, E. and Haselschwardt, S. J. and Hernandez, M. A. and Hertel, S. A. and Heuermann, G. and Homenides, G. J. and Horn, M. and Huang, D. Q. and Hunt, D. and Jacquet, E. and James, R. S. and Johnson, J. and Kaboth, A. C. and Kamaha, A. C. and K., Meghna K. and Khaitan, D. and Khazov, A. and Khurana, I. and Kim, J. and Kim, Y. D. and Kingston, J. and Kirk, R. and Kodroff, D. and Korley, L. and Korolkova, E. V. and Kraus, H. and Kravitz, S. and Kreczko, L. and Kudryavtsev, V. A. and Lawes, C. and Leonard, D. S. and Lesko, K. T. and Levy, C. and Lin, J. and Lindote, A. and Lippincott, W. H. and Lopes, M. I. and Lorenzon, W. and Lu, C. and Luitz, S. and Majewski, P. A. and Manalaysay, A. and Mannino, R. L. and Maupin, C. and McCarthy, M. E. and McDowell, G. and McKinsey, D. N. and McLaughlin, J. and McLaughlin, J. B. and McMonigle, R. and Mizrachi, E. and Monte, A. and Monzani, M. E. and Mendoza, J. D. Morales and Morrison, E. and Mount, B. J. and Murdy, M. and Murphy, A. St. J. and Naylor, A. and Nelson, H. N. and Neves, F. and Nguyen, A. and O’Brien, C. L. and Olcina, I. and Oliver-Mallory, K. C. and Orpwood, J. and Oyulmaz, K. Y and Palladino, K. J. and Palmer, J. and Pannifer, N. J. and Parveen, N. and Patton, S. J. and Penning, B. and Pereira, G. and Perry, E. and Pershing, T. and Piepke, A. and Qie, Y. and Reichenbacher, J. and Rhyne, C. A. and Richards, A. and Riffard, Q. and Rischbieter, G. R. C. and Ritchey, E. and Riyat, H. S. and Rosero, R. and Rushton, T. and Rynders, D. and Santone, D. and Sazzad, A. B. M. R. and Schnee, R. W. and Sehr, G. and Shafer, B. and Shaw, S. and Shutt, T. and Silk, J. J. and Silva, C. and Sinev, G. and Siniscalco, J. and Smith, R. and Solovov, V. N. and Sorensen, P. and Soria, J. and Stancu, I. and Stevens, A. and Stifter, K. and Suerfu, B. and Sumner, T. J. and Szydagis, M. and Tiedt, D. R. and Timalsina, M. and Tong, Z. and Tovey, D. R. and Tranter, J. and Trask, M. and Tripathi, M. and Usón, A. and Vacheret, A. and Vaitkus, A. C. and Valentino, O. and Velan, V. and Wang, A. and Wang, J. J. and Wang, Y. and Watson, J. R. and Weeldreyer, L. and Whitis, T. J. and Wild, K. and Williams, M. and Wisniewski, W. J. and Wolf, L. and Wolfs, F. L. H. and Woodford, S. and Woodward, D. and Wright, C. J. and Xia, Q. and Xu, J. and Xu, Y. and Yeh, M. and Yeum, D. and Zha, W. and Zweig, E. A. and },
   year={2025},
   month=July }

@article{Aprile_2025,
   title={WIMP Dark Matter Search Using a 3.1 Tonne-Year Exposure of the XENONnT Experiment},
   volume={135},
   ISSN={1079-7114},
   url={http://dx.doi.org/10.1103/msw4-t342},
   DOI={10.1103/msw4-t342},
   number={22},
   journal={Physical Review Letters},
   publisher={American Physical Society (APS)},
   author={Aprile, E. and Aalbers, J. and Abe, K. and Ahmed Maouloud, S. and Althueser, L. and Andrieu, B. and Angelino, E. and Antón Martin, D. and Armbruster, S. R. and Arneodo, F. and Baudis, L. and Bazyk, M. and Bellagamba, L. and Biondi, R. and Bismark, A. and Boese, K. and Brown, A. and Bruno, G. and Budnik, R. and Cai, C. and Capelli, C. and Cardoso, J. M. R. and Cimental Chávez, A. P. and Colijn, A. P. and Conrad, J. and Cuenca-García, J. J. and D’Andrea, V. and Daniel Garcia, L. C. and Decowski, M. P. and Deisting, A. and Di Donato, C. and Di Gangi, P. and Diglio, S. and Eitel, K. and el Morabit, S. and Elykov, A. and Ferella, A. D. and Ferrari, C. and Fischer, H. and Flehmke, T. and Flierman, M. and Fuchs, D. and Fulgione, W. and Fuselli, C. and Gaemers, P. and Gaior, R. and Gao, F. and Ghosh, S. and Giacomobono, R. and Girard, F. and Glade-Beucke, R. and Grandi, L. and Grigat, J. and Guan, H. and Guida, M. and Gyorgy, P. and Hammann, R. and Higuera, A. and Hils, C. and Hoetzsch, L. and Hood, N. F. and Iacovacci, M. and Itow, Y. and Jakob, J. and Joerg, F. and Kaminaga, Y. and Kara, M. and Kavrigin, P. and Kazama, S. and Kharbanda, P. and Kobayashi, M. and Koke, D. and Kooshkjalali, K. and Kopec, A. and Landsman, H. and Lang, R. F. and Levinson, L. and Li, I. and Li, S. and Liang, S. and Liang, Z. and Lin, Y.-T. and Lindemann, S. and Lindner, M. and Liu, K. and Liu, M. and Loizeau, J. and Lombardi, F. and Long, J. and Lopes, J. A. M. and Lucchetti, G. M. and Luce, T. and Ma, Y. and Macolino, C. and Mahlstedt, J. and Mancuso, A. and Manenti, L. and Marignetti, F. and Marrodán Undagoitia, T. and Martens, K. and Masbou, J. and Mastroianni, S. and Melchiorre, A. and Merz, J. and Messina, M. and Michael, A. and Miuchi, K. and Molinario, A. and Moriyama, S. and Morå, K. and Mosbacher, Y. and Murra, M. and Müller, J. and Ni, K. and Oberlack, U. and Paetsch, B. and Pan, Y. and Pellegrini, Q. and Peres, R. and Peters, C. and Pienaar, J. and Pierre, M. and Plante, G. and Pollmann, T. R. and Principe, L. and Qi, J. and Qin, J. and Ramírez García, D. and Rajado, M. and Ravindran, A. and Razeto, A. and Singh, R. and Sanchez, L. and dos Santos, J. M. F. and Sarnoff, I. and Sartorelli, G. and Schreiner, J. and Schulte, P. and Schulze Eißing, H. and Schumann, M. and Scotto Lavina, L. and Selvi, M. and Semeria, F. and Shagin, P. and Shi, S. and Shi, J. and Silva, M. and Simgen, H. and Stevens, A. and Szyszka, C. and Takeda, A. and Takeuchi, Y. and Tan, P.-L. and Thers, D. and Trinchero, G. and Tunnell, C. D. and Tönnies, F. and Valerius, K. and Vecchi, S. and Vetter, S. and Villazon Solar, F. I. and Volta, G. and Weinheimer, C. and Weiss, M. and Wenz, D. and Wittweg, C. and Wu, V. H. S. and Xing, Y. and Xu, D. and Xu, Z. and Yamashita, M. and Yang, J. and Yang, L. and Ye, J. and Yuan, L. and Zavattini, G. and Zhao, Y. and Zhong, M. and },
   year={2025},
   month=Nov }

@article{Bo_2025,
   title={Dark Matter Search Results from 1.54 Tonne-Year Exposure of {PandaX-4T}},
   volume={134},
   ISSN={1079-7114},
   url={http://dx.doi.org/10.1103/PhysRevLett.134.011805},
   DOI={10.1103/physrevlett.134.011805},
   number={1},
   journal={Physical Review Letters},
   publisher={American Physical Society (APS)},
   author={Bo, Zihao and Chen, Wei and Chen, Xun and Chen, Yunhua and Cheng, Zhaokan and Cui, Xiangyi and Fan, Yingjie and Fang, Deqing and Gao, Zhixing and Geng, Lisheng and Giboni, Karl and Guo, Xunan and Guo, Xuyuan and Guo, Zichao and Han, Chencheng and Han, Ke and He, Changda and He, Jinrong and Huang, Di and Huang, Houqi and Huang, Junting and Hou, Ruquan and Hou, Yu and Ji, Xiangdong and Ji, Xiangpan and Ju, Yonglin and Li, Chenxiang and Li, Jiafu and Li, Mingchuan and Li, Shuaijie and Li, Tao and Li, Zhiyuan and Lin, Qing and Liu, Jianglai and Lu, Congcong and Lu, Xiaoying and Luo, Lingyin and Luo, Yunyang and Ma, Wenbo and Ma, Yugang and Mao, Yajun and Meng, Yue and Ning, Xuyang and Pang, Binyu and Qi, Ningchun and Qian, Zhicheng and Ren, Xiangxiang and Shan, Dong and Shang, Xiaofeng and Shao, Xiyuan and Shen, Guofang and Shen, Manbin and Sun, Wenliang and Tao, Yi and Wang, Anqing and Wang, Guanbo and Wang, Hao and Wang, Jiamin and Wang, Lei and Wang, Meng and Wang, Qiuhong and Wang, Shaobo and Wang, Siguang and Wang, Wei and Wang, Xiuli and Wang, Xu and Wang, Zhou and Wei, Yuehuan and Wu, Weihao and Wu, Yuan and Xiao, Mengjiao and Xiao, Xiang and Xiong, Kaizhi and Xu, Yifan and Yao, Shunyu and Yan, Binbin and Yan, Xiyu and Yang, Yong and Ye, Peihua and Yu, Chunxu and Yuan, Ying and Yuan, Zhe and Yun, Youhui and Zeng, Xinning and Zhang, Minzhen and Zhang, Peng and Zhang, Shibo and Zhang, Shu and Zhang, Tao and Zhang, Wei and Zhang, Yang and Zhang, Yingxin and Zhang, Yuanyuan and Zhao, Li and Zhou, Jifang and Zhou, Jiaxu and Zhou, Jiayi and Zhou, Ning and Zhou, Xiaopeng and Zhou, Yubo and Zhou, Zhizhen and },
   year={2025},
   month=Jan }

@misc{essig2016directdetectionsubgevdark,
      title={Direct Detection of sub-GeV Dark Matter with Semiconductor Targets}, 
      author={Rouven Essig and Marivi Fernandez-Serra and Jeremy Mardon and Adrian Soto and Tomer Volansky and Tien-Tien Yu},
      year={2016},
      eprint={1509.01598},
      archivePrefix={arXiv},
      primaryClass={hep-ph},
      url={https://arxiv.org/abs/1509.01598}, 
}

@article{Hochberg_2016,
   title={Detecting ultralight bosonic dark matter via absorption in superconductors},
   volume={94},
   ISSN={2470-0029},
   url={http://dx.doi.org/10.1103/PhysRevD.94.015019},
   DOI={10.1103/physrevd.94.015019},
   number={1},
   journal={Physical Review D},
   publisher={American Physical Society (APS)},
   author={Hochberg, Yonit and Lin, Tongyan and Zurek, Kathryn M.},
   year={2016},
   month=July }

@article{PhysRevLett.123.151802,
  title = {Detecting Sub-GeV Dark Matter with Superconducting Nanowires},
  author = {Hochberg, Yonit and Charaev, Ilya and Nam, Sae-Woo and Verma, Varun and Colangelo, Marco and Berggren, Karl K.},
  journal = {Phys. Rev. Lett.},
  volume = {123},
  issue = {15},
  pages = {151802},
  numpages = {6},
  year = {2019},
  month = {Oct},
  publisher = {American Physical Society},
  doi = {10.1103/PhysRevLett.123.151802},
  url = {https://link.aps.org/doi/10.1103/PhysRevLett.123.151802}
}

@article{Knapen_2018,
   title={Detection of light dark matter with optical phonons in polar materials},
   volume={785},
   ISSN={0370-2693},
   url={http://dx.doi.org/10.1016/j.physletb.2018.08.064},
   DOI={10.1016/j.physletb.2018.08.064},
   journal={Physics Letters B},
   publisher={Elsevier BV},
   author={Knapen, Simon and Lin, Tongyan and Pyle, Matt and Zurek, Kathryn M.},
   year={2018},
   month=Oct, pages={386–390} }

@article{Baudis_2025,
   title={First Sub-MeV Dark Matter Search with the QROCODILE Experiment Using Superconducting Nanowire Single-Photon Detectors},
   volume={135},
   ISSN={1079-7114},
   url={http://dx.doi.org/10.1103/4hb6-f6jl},
   DOI={10.1103/4hb6-f6jl},
   number={8},
   journal={Physical Review Letters},
   publisher={American Physical Society (APS)},
   author={Baudis, Laura and Bismark, Alexander and Brugger, Noah and Capelli, Chiara and Charaev, Ilya and García, Jose Cuenca and Hadas, Guy Daniel and Hochberg, Yonit and Hohmann, Judith K. and Kavner, Alexander and Koos, Christian and Kuzmin, Artem and Lehmann, Benjamin V. and Nägeli, Severin and Neupert, Titus and Penning, Bjoern and García, Diego Ramírez and Schilling, Andreas},
   year={2025},
   month=Aug }

@article{Hochberg_2018,
   title={Detection of sub-MeV dark matter with three-dimensional Dirac materials},
   volume={97},
   ISSN={2470-0029},
   url={http://dx.doi.org/10.1103/PhysRevD.97.015004},
   DOI={10.1103/physrevd.97.015004},
   number={1},
   journal={Physical Review D},
   publisher={American Physical Society (APS)},
   author={Hochberg, Yonit and Kahn, Yonatan and Lisanti, Mariangela and Zurek, Kathryn M. and Grushin, Adolfo G. and Ilan, Roni and Griffin, Sinéad M. and Liu, Zhen-Fei and Weber, Sophie F. and Neaton, Jeffrey B.},
   year={2018},
   month=Jan }

@article{Geilhufe_2020,
   title={Dirac materials for sub-MeV dark matter detection: New targets and improved formalism},
   volume={101},
   ISSN={2470-0029},
   url={http://dx.doi.org/10.1103/PhysRevD.101.055005},
   DOI={10.1103/physrevd.101.055005},
   number={5},
   journal={Physical Review D},
   publisher={American Physical Society (APS)},
   author={Geilhufe, R. Matthias and Kahlhoefer, Felix and Winkler, Martin Wolfgang},
   year={2020},
   month=Mar }

@misc{das2024submevdarkmatterdetection,
      title={Sub-MeV Dark Matter Detection with Bilayer Graphene}, 
      author={Anirban Das and Jiho Jang and Hongki Min},
      year={2024},
      eprint={2312.00866},
      archivePrefix={arXiv},
      primaryClass={hep-ph},
      url={https://arxiv.org/abs/2312.00866}, 
}

@misc{sherpa2026divedeepersubmarinesubmev,
      title={Dive deeper with SUBMARINE: SUB-Mev dArk matter diRect detectIon using bilayer grapheNE}, 
      author={Rinchen Sherpa and Anuvab Sarkar and Tarak Nath Maity and Paramita Dutta and Ranjan Laha and Anirban Das},
      year={2026},
      eprint={2604.21969},
      archivePrefix={arXiv},
      primaryClass={hep-ph},
      url={https://arxiv.org/abs/2604.21969}, 
}

@article{PhysRevLett.42.407,
  title = {Dynamical Role of Light Neutral Leptons in Cosmology},
  author = {Tremaine, Scott and Gunn, James E.},
  journal = {Phys. Rev. Lett.},
  volume = {42},
  issue = {6},
  pages = {407--410},
  numpages = {0},
  year = {1979},
  month = {Feb},
  publisher = {American Physical Society},
  doi = {10.1103/PhysRevLett.42.407},
  url = {https://link.aps.org/doi/10.1103/PhysRevLett.42.407}
}

@article{33eb53d663e34ecc82e4e64ea6ca5cbe,
title = "'Luttinger liquid theory' of one-dimensional quantum fluids. I. Properties of the Luttinger model and their extension to the general 1D interacting spinless fermi gas",
author = "Haldane, \{F. D.M.\}",
year = "1981",
doi = "10.1088/0022-3719/14/19/010",
volume = "14",
pages = "2585--2609",
journal = "Journal of Physics C: Solid State Physics",
issn = "0022-3719",
publisher = "Institute of Physics",
number = "19",
}

@article{Voit_1995,
   title={One-dimensional Fermi liquids},
   volume={58},
   ISSN={1361-6633},
   url={http://dx.doi.org/10.1088/0034-4885/58/9/002},
   DOI={10.1088/0034-4885/58/9/002},
   number={9},
   journal={Reports on Progress in Physics},
   publisher={IOP Publishing},
   author={Voit, J},
   year={1995},
   month=Sept, pages={977–1116} }

@article{PhysRevLett.79.5086,
  title = {Coulomb Interactions and Mesoscopic Effects in Carbon Nanotubes},
  author = {Kane, Charles and Balents, Leon and Fisher, Matthew P. A.},
  journal = {Phys. Rev. Lett.},
  volume = {79},
  issue = {25},
  pages = {5086--5089},
  numpages = {0},
  year = {1997},
  month = {Dec},
  publisher = {American Physical Society},
  doi = {10.1103/PhysRevLett.79.5086},
  url = {https://link.aps.org/doi/10.1103/PhysRevLett.79.5086}
}

@article{Egger_1997,
   title={Effective Low-Energy Theory for Correlated Carbon Nanotubes},
   volume={79},
   ISSN={1079-7114},
   url={http://dx.doi.org/10.1103/PhysRevLett.79.5082},
   DOI={10.1103/physrevlett.79.5082},
   number={25},
   journal={Physical Review Letters},
   publisher={American Physical Society (APS)},
   author={Egger, Reinhold and Gogolin, Alexander O.},
   year={1997},
   month=Dec, pages={5082–5085} }

@article{Egger_1998,
   title={Correlated transport and non-Fermi-liquid behavior in single-wall carbon nanotubes},
   volume={3},
   ISSN={1434-6028},
   url={http://dx.doi.org/10.1007/s100510050315},
   DOI={10.1007/s100510050315},
   number={3},
   journal={The European Physical Journal B},
   publisher={Springer Science and Business Media LLC},
   author={Egger, R. and Gogolin, A.O.},
   year={1998},
   month=July, pages={281–300} }

@article{Imambekov_2012,
   title={One-dimensional quantum liquids: Beyond the Luttinger liquid paradigm},
   volume={84},
   ISSN={1539-0756},
   url={http://dx.doi.org/10.1103/RevModPhys.84.1253},
   DOI={10.1103/revmodphys.84.1253},
   number={3},
   journal={Reviews of Modern Physics},
   publisher={American Physical Society (APS)},
   author={Imambekov, Adilet and Schmidt, Thomas L. and Glazman, Leonid I.},
   year={2012},
   month=Sept, pages={1253–1306} }

@article{Shi_2015,
author = {Shi, Zhiwen and Hong, Xiaoping and Bechtel, Hans and Zeng, Bo and Martin, Michael and Watanabe, Kenji and Taniguchi, Takashi and Shen, Yuen-Ron and Wang, Feng},
year = {2015},
month = {07},
pages = {},
title = {Observation of a Luttinger-liquid plasmon in metallic single-walled carbon nanotubes},
volume = {9},
journal = {Nature Photonics},
doi = {10.1038/nphoton.2015.123}
}

@misc{egger2000luttingerliquidbehaviormetallic,
      title={Luttinger liquid behavior in metallic carbon nanotubes}, 
      author={R. Egger and A. Bachtold and M. Fuhrer and M. Bockrath and D. Cobden and P. McEuen},
      year={2000},
      eprint={cond-mat/0008008},
      archivePrefix={arXiv},
      primaryClass={cond-mat.mes-hall},
      url={https://arxiv.org/abs/cond-mat/0008008}, 
}

@article{Ishii_2004,
author = {Ishii, Hiroyoshi and Kataura, Hiromichi and Shiozawa, Hidetsugu and Yoshioka, Hideo and Otsubo, Hideo and Takayama, Yasuhiro and Miyahara, Tsuneaki and Suzuki, Shinzo and Achiba, Yohji and Nakatake, Masashi and Narimura, Takamasa and Higashiguchi, Mitsuharu and Shimada, Kazuyuki and Namatame, Hiroaki and Taniguchi, Masaki},
year = {2004},
month = {01},
pages = {540-4},
title = {Direct observation of Tomonaga-Luttinger-liquid state in carbon nanotubes at low temperatures},
volume = {426},
journal = {Nature},
doi = {10.1038/nature02074}
}

@article{PhysRevD.103.016006,
  title = {Directional dark matter detection in anisotropic Dirac materials},
  author = {Coskuner, Ahmet and Mitridate, Andrea and Olivares, Andres and Zurek, Kathryn M.},
  journal = {Phys. Rev. D},
  volume = {103},
  issue = {1},
  pages = {016006},
  numpages = {12},
  year = {2021},
  month = {Jan},
  publisher = {American Physical Society},
  doi = {10.1103/PhysRevD.103.016006},
  url = {https://link.aps.org/doi/10.1103/PhysRevD.103.016006}
}

@article{PhysRevD.108.015015,
  title = {Directional detection of dark matter with anisotropic response functions},
  author = {Boyd, Christian and Hochberg, Yonit and Kahn, Yonatan and Kramer, Eric David and Kurinsky, Noah and Lehmann, Benjamin V. and Yu, To Chin},
  journal = {Phys. Rev. D},
  volume = {108},
  issue = {1},
  pages = {015015},
  numpages = {20},
  year = {2023},
  month = {Jul},
  publisher = {American Physical Society},
  doi = {10.1103/PhysRevD.108.015015},
  url = {https://link.aps.org/doi/10.1103/PhysRevD.108.015015}
}

@article{Zhang_2011,
author = {Zhang, Qiang and Huang, Jia-Qi and Zhao, Meng-Qiang and Qian, Wei-Zhong and Wei, Fei},
title = {Carbon Nanotube Mass Production: Principles and Processes},
journal = {ChemSusChem},
volume = {4},
number = {7},
pages = {864-889},
doi = {https://doi.org/10.1002/cssc.201100177},
url = {https://chemistry-europe.onlinelibrary.wiley.com/doi/abs/10.1002/cssc.201100177},
eprint = {https://chemistry-europe.onlinelibrary.wiley.com/doi/pdf/10.1002/cssc.201100177},
year = {2011}
}

@article{hughes_2024,
author = {Hughes, Kevin and Iyer, Kavita and Bird, Robert and Ivanov, Julian and Banerjee, Saswata and Georges, Gilles and Zhou, Qiongqiong},
year = {2024},
month = {08},
pages = {},
title = {Review of Carbon Nanotube Research and Development: Materials and Emerging Applications},
volume = {7},
journal = {ACS Applied Nano Materials},
doi = {10.1021/acsanm.4c02721}
}

@article{Walker_2019,
   title={Global Alignment of Solution-Based Single-Wall Carbon Nanotube Films via Machine-Vision Controlled Filtration},
   volume={19},
   ISSN={1530-6992},
   url={http://dx.doi.org/10.1021/acs.nanolett.9b02853},
   DOI={10.1021/acs.nanolett.9b02853},
   number={10},
   journal={Nano Letters},
   publisher={American Chemical Society (ACS)},
   author={Walker, Joshua S. and Fagan, Jeffrey A. and Biacchi, Adam J. and Kuehl, Valerie A. and Searles, Thomas A. and Hight Walker, Angela R. and Rice, William D.},
   year={2019},
   month=Sept, pages={7256–7264} }

@ARTICLE{2018NatAs...2...90E,
       author = {{Echternach}, P.~M. and {Pepper}, B.~J. and {Reck}, T. and {Bradford}, C.~M.},
        title = "{Single photon detection of 1.5 THz radiation with the quantum capacitance detector}",
      journal = {Nature Astronomy},
         year = 2018,
        month = nov,
       volume = {2},
        pages = {90-97},
          doi = {10.1038/s41550-017-0294-y},
       adsurl = {https://ui.adsabs.harvard.edu/abs/2018NatAs...2...90E}
}

@article{Fink_2024,
   title={Superconducting quasiparticle-amplifying transmon: A qubit-based sensor for meV-scale phonons and single terahertz photons},
   volume={22},
   ISSN={2331-7019},
   url={http://dx.doi.org/10.1103/PhysRevApplied.22.054009},
   DOI={10.1103/physrevapplied.22.054009},
   number={5},
   journal={Physical Review Applied},
   publisher={American Physical Society (APS)},
   author={Fink, C.W. and Salemi, C.P. and Young, B.A. and Schuster, D.I. and Kurinsky, N.A.},
   year={2024},
   month=Nov }

@article{PhysRevLett.127.111301,
  title = {Cosmology of Sub-MeV Dark Matter Freeze-In},
  author = {Dvorkin, Cora and Lin, Tongyan and Schutz, Katelin},
  journal = {Phys. Rev. Lett.},
  volume = {127},
  issue = {11},
  pages = {111301},
  numpages = {7},
  year = {2021},
  month = {Sep},
  publisher = {American Physical Society},
  doi = {10.1103/PhysRevLett.127.111301},
  url = {https://link.aps.org/doi/10.1103/PhysRevLett.127.111301}
}

@article{Knapen_2021,
   title={Dark matter-electron scattering in dielectrics},
   volume={104},
   ISSN={2470-0029},
   url={http://dx.doi.org/10.1103/PhysRevD.104.015031},
   DOI={10.1103/physrevd.104.015031},
   number={1},
   journal={Physical Review D},
   publisher={American Physical Society (APS)},
   author={Knapen, Simon and Kozaczuk, Jonathan and Lin, Tongyan},
   year={2021},
   month=July }

@article{Hochberg_2021,
   title={Determining Dark-Matter–Electron Scattering Rates from the Dielectric Function},
   volume={127},
   ISSN={1079-7114},
   url={http://dx.doi.org/10.1103/PhysRevLett.127.151802},
   DOI={10.1103/physrevlett.127.151802},
   number={15},
   journal={Physical Review Letters},
   publisher={American Physical Society (APS)},
   author={Hochberg, Yonit and Kahn, Yonatan and Kurinsky, Noah and Lehmann, Benjamin V. and Yu, To Chin and Berggren, Karl K.},
   year={2021},
   month=Oct }

@article{Trickle_2020,
   title={Multi-channel direct detection of light dark matter: theoretical framework},
   volume={2020},
   ISSN={1029-8479},
   url={http://dx.doi.org/10.1007/JHEP03(2020)036},
   DOI={10.1007/jhep03(2020)036},
   number={3},
   journal={Journal of High Energy Physics},
   publisher={Springer Science and Business Media LLC},
   author={Trickle, Tanner and Zhang, Zhengkang and Zurek, Kathryn M. and Inzani, Katherine and Griffin, Sinéad M.},
   year={2020},
   month=Mar }

@article{Hamley_2023,
author = {Hamley, Ian and Castelletto, Valeria},
year = {2023},
month = {07},
pages = {102959},
title = {Small-angle scattering techniques for peptide and peptide hybrid nanostructures and peptide-based biomaterials},
volume = {318},
journal = {Advances in Colloid and Interface Science},
doi = {10.1016/j.cis.2023.102959}
}

@article{PhysRevB.64.241402,
  title = {Thermal expansion of single-walled carbon nanotube (SWNT) bundles: X-ray diffraction studies},
  author = {Maniwa, Yutaka and Fujiwara, Ryuji and Kira, Hiroshi and Tou, Hideki and Kataura, Hiromichi and Suzuki, Shinzo and Achiba, Yohji and Nishibori, Eiji and Takata, Masaki and Sakata, Makoto and Fujiwara, Akihiko and Suematsu, Hiroyoshi},
  journal = {Phys. Rev. B},
  volume = {64},
  issue = {24},
  pages = {241402(R)},
  numpages = {3},
  year = {2001},
  month = {Nov},
  publisher = {American Physical Society},
  doi = {10.1103/PhysRevB.64.241402},
  url = {https://link.aps.org/doi/10.1103/PhysRevB.64.241402}
}

@ARTICLE{1996APh.....6...87L,
       author = {{Lewin}, J.~D. and {Smith}, P.~F.},
        title = "{Review of mathematics, numerical factors, and corrections for dark matter experiments based on elastic nuclear recoil}",
      journal = {Astroparticle Physics},
         year = 1996,
        month = dec,
       volume = {6},
       number = {1},
        pages = {87-112},
          doi = {10.1016/S0927-6505(96)00047-3},
       adsurl = {https://ui.adsabs.harvard.edu/abs/1996APh.....6...87L}
}

@article{PhysRevD.37.3388,
  title = {Signal modulation in cold-dark-matter detection},
  author = {Freese, Katherine and Frieman, Joshua and Gould, Andrew},
  journal = {Phys. Rev. D},
  volume = {37},
  issue = {12},
  pages = {3388--3405},
  numpages = {0},
  year = {1988},
  month = {Jun},
  publisher = {American Physical Society},
  doi = {10.1103/PhysRevD.37.3388},
  url = {https://link.aps.org/doi/10.1103/PhysRevD.37.3388}
}

@article{Smith_2007,
   title={The RAVE survey: constraining the local Galactic escape speed},
   volume={379},
   ISSN={1365-2966},
   url={http://dx.doi.org/10.1111/j.1365-2966.2007.11964.x},
   DOI={10.1111/j.1365-2966.2007.11964.x},
   number={2},
   journal={Monthly Notices of the Royal Astronomical Society},
   publisher={Oxford University Press (OUP)},
   author={Smith, M. C. and Ruchti, G. R. and Helmi, A. and Wyse, R. F. G. and Fulbright, J. P. and Freeman, K. C. and Navarro, J. F. and Seabroke, G. M. and Steinmetz, M. and Williams, M. and Bienayme, O. and Binney, J. and Bland-Hawthorn, J. and Dehnen, W. and Gibson, B. K. and Gilmore, G. and Grebel, E. K. and Munari, U. and Parker, Q. A. and Scholz, R.- D. and Siebert, A. and Watson, F. G. and Zwitter, T.},
   year={2007},
   month=Aug, pages={755–772} }

@article{Bockrath_1999,
   title={Luttinger-liquid behaviour in carbon nanotubes},
   volume={397},
   ISSN={1476-4687},
   url={http://dx.doi.org/10.1038/17569},
   DOI={10.1038/17569},
   number={6720},
   journal={Nature},
   publisher={Springer Science and Business Media LLC},
   author={Bockrath, Marc and Cobden, David H. and Lu, Jia and Rinzler, Andrew G. and Smalley, Richard E. and Balents, Leon and McEuen, Paul L.},
   year={1999},
   month=Feb, pages={598–601} }

@INPROCEEDINGS{2013APS..MAR.N8001C,
       author = {{Chudow}, Joel D. and {McKitterick}, Chris B. and {Prober}, Daniel E. and {Santavicca}, Daniel F. and {Kim}, Philip},
        title = "{Terahertz Detection as a Probe of Luttinger-Liquid Behavior in an Individual Single-Walled Carbon Nanotube}",
    booktitle = {APS March Meeting Abstracts},
         year = 2013,
       series = {APS Meeting Abstracts},
       volume = {2013},
        month = mar,
          eid = {N8.001},
        pages = {N8.001},
       adsurl = {https://ui.adsabs.harvard.edu/abs/2013APS..MAR.N8001C}
}

@article{Riccardo_Catena_2010,
   title={A novel determination of the local dark matter density},
   volume={2010},
   ISSN={1475-7516},
   url={http://dx.doi.org/10.1088/1475-7516/2010/08/004},
   DOI={10.1088/1475-7516/2010/08/004},
   number={08},
   journal={Journal of Cosmology and Astroparticle Physics},
   publisher={IOP Publishing},
   author={Riccardo Catena and Piero Ullio},
   year={2010},
   month=Aug, pages={004–004} }

@article{de_Salas_2021,
   title={Dark matter local density determination: recent observations and future prospects},
   volume={84},
   ISSN={1361-6633},
   url={http://dx.doi.org/10.1088/1361-6633/ac24e7},
   DOI={10.1088/1361-6633/ac24e7},
   number={10},
   journal={Reports on Progress in Physics},
   publisher={IOP Publishing},
   author={de Salas, Pablo F and Widmark, A},
   year={2021},
   month=Oct, pages={104901} }

@article{Ranjan_2015,
   title={Clean carbon nanotubes coupled to superconducting impedance-matching circuits},
   volume={6},
   ISSN={2041-1723},
   url={http://dx.doi.org/10.1038/ncomms8165},
   DOI={10.1038/ncomms8165},
   number={1},
   journal={Nature Communications},
   publisher={Springer Science and Business Media LLC},
   author={Ranjan, V. and Puebla-Hellmann, G. and Jung, M. and Hasler, T. and Nunnenkamp, A. and Muoth, M. and Hierold, C. and Wallraff, A. and Schönenberger, C.},
   year={2015},
   month=May }

@article{PhysRevApplied.15.064050,
  title = {Circuit Quantum Electrodynamics with Carbon-Nanotube-Based Superconducting Quantum Circuits},
  author = {Mergenthaler, Matthias and Nersisyan, Ani and Patterson, Andrew and Esposito, Martina and Baumgartner, Andreas and Sch\"onenberger, Christian and Briggs, G. Andrew D. and Laird, Edward A. and Leek, Peter J.},
  journal = {Phys. Rev. Appl.},
  volume = {15},
  issue = {6},
  pages = {064050},
  numpages = {8},
  year = {2021},
  month = {Jun},
  publisher = {American Physical Society},
  doi = {10.1103/PhysRevApplied.15.064050},
  url = {https://link.aps.org/doi/10.1103/PhysRevApplied.15.064050}
}

@misc{magoon2026demonstrationsquatdetectorarchitecture,
      title={A First Demonstration of the SQUAT Detector Architecture: Direct Measurement of Resonator-Free Charge-Sensitive Transmons}, 
      author={H. Magoon and T. Aralis and T. Dyson and J. Anczarski and D. Baxter and G. Bratrud and R. Carpenter and S. Condon and A. Droster and E. Figueroa-Feliciano and C. W. Fink and S. Harvey and A. Simchony and Z. J. Smith and S. Stevens and N. Tabassum and B. A. Young and C. P. Salemi and K. Stifter and D. I. Schuster and N. A. Kurinsky},
      year={2026},
      eprint={2601.16261},
      archivePrefix={arXiv},
      primaryClass={physics.ins-det},
      url={https://arxiv.org/abs/2601.16261}, 
}

@misc{ramanathan2026quantumparitydetectorsqubit,
      title={Quantum Parity Detectors: a qubit based particle detection scheme with meV thresholds for rare-event searches}, 
      author={Karthik Ramanathan and Brandon J. Sandoval and John E. Parker and Lalit M. Joshi and Andrew D. Beyer and Pierre M. Echternach and Serge Rosenblum and Sunil R. Golwala},
      year={2026},
      eprint={2405.17192},
      archivePrefix={arXiv},
      primaryClass={physics.ins-det},
      doi={https://doi.org/10.1103/kqd2-spb1},
      url={https://arxiv.org/abs/2405.17192}, 
}

@misc{NanoIntegris_MetallicSWNTs,
  author       = {{NanoIntegris}},
  title        = {{Metallic SWNTs (IsoNanotubes-M)}},
  howpublished = {\url{https://nanointegris.com/our-products/metallic-swnts-isonanotubes-m/}},
  note         = {Accessed: 2026-06-11}
}

@misc{NanoIntegris_Catalog_2020,
  author       = {{NanoIntegris}},
  title        = {{2020 Product Catalog}},
  year         = {2020},
  howpublished = {\url{https://nanointegris.com/wp-content/uploads/2020/03/NanoIntegrisCatalog_2020.pdf}},
  note         = {Accessed: 2026-06-11}
}

@misc{SigmaAldrich_IsoNanotubesM_750530,
  author       = {{Sigma-Aldrich}},
  title        = {{Carbon nanotube, single-walled, 98\% metallic, product no. 750530}},
  howpublished = {\url{https://www.sigmaaldrich.com/US/en/product/aldrich/750530}},
  note         = {Accessed: 2026-06-11}
}

@misc{Raymor_IsoNanotubesM,
  author       = {{Raymor Industries}},
  title        = {{Metallic Single-Wall Carbon Nanotubes / m-SWCNT}},
  howpublished = {\url{https://raymor.com/our-products/isonanotubes-m/}},
  note         = {Accessed: 2026-06-11}
}

@misc{Graphenea_GrapheneProperties,
  author       = {{Graphenea}},
  title        = {{Properties of Graphene}},
  howpublished = {\url{https://www.graphenea.com/pages/graphene-properties}},
  note         = {Accessed: 2026-06-11}
}

@article{Li_2025,
   title={Cosmic-ray-induced correlated errors in superconducting qubit array},
   volume={16},
   ISSN={2041-1723},
   url={http://dx.doi.org/10.1038/s41467-025-59778-z},
   DOI={10.1038/s41467-025-59778-z},
   number={1},
   journal={Nature Communications},
   publisher={Springer Science and Business Media LLC},
   author={Li, Xuegang and Wang, Junhua and Jiang, Yao-Yao and Xue, Guang-Ming and Cai, Xiaoxia and Zhou, Jun and Gong, Ming and Liu, Zhao-Feng and Zheng, Shuang-Yu and Ma, Deng-Ke and Chen, Mo and Sun, Wei-Jie and Yang, Shuang and Yan, Fei and Jin, Yi-Rong and Zhao, S. P. and Ding, Xue-Feng and Yu, Hai-Feng},
   year={2025},
   month=May }

@article{PhysRevD.73.053004,
  title = {Muon-induced background study for underground laboratories},
  author = {Mei, D.-M. and Hime, A.},
  journal = {Phys. Rev. D},
  volume = {73},
  issue = {5},
  pages = {053004},
  numpages = {18},
  year = {2006},
  month = {Mar},
  publisher = {American Physical Society},
  doi = {10.1103/PhysRevD.73.053004},
  url = {https://link.aps.org/doi/10.1103/PhysRevD.73.053004}
}

@misc{akerib2026measurementmuonfluxsanford,
      title={Measurement of the Muon Flux at the Sanford Underground Research Facility with the LUX-ZEPLIN Dark Matter Detector}, 
      author={D. S. Akerib and A. K. Al Musalhi and F. Alder and B. J. Almquist and C. S. Amarasinghe and A. Ames and T. J. Anderson and N. Angelides and H. M. Araújo and J. E. Armstrong and M. Arthurs and A. Baker and S. Balashov and J. Bang and J. W. Bargemann and E. E. Barillier and K. Beattie and A. Bhatti and T. P. Biesiadzinski and H. J. Birch and E. Bishop and G. M. Blockinger and C. A. J. Brew and P. Brás and S. Burdin and M. C. Carmona-Benitez and M. Carter and A. Chawla and H. Chen and Y. T. Chin and N. I. Chott and S. Contreras and M. V. Converse and R. Coronel and A. Cottle and G. Cox and D. Curran and C. E. Dahl and I. Darlington and S. Dave and A. David and J. Delgaudio and S. Dey and L. de Viveiros and L. Di Felice and C. Ding and J. E. Y. Dobson and E. Druszkiewicz and S. Dubey and C. L. Dunbar and S. R. Eriksen and N. M. Fearon and N. Fieldhouse and S. Fiorucci and H. Flaecher and E. D. Fraser and T. M. A. Fruth and P. W. Gaemers and R. J. Gaitskell and A. Geffre and J. Genovesi and C. Ghag and J. Ghamsari and A. Ghosh and R. Gibbons and S. Gokhale and J. Green and M. G. D. van der Grinten and J. J. Haiston and C. R. Hall and T. Hall and R. H. Hampp and S. J. Haselschwardt and M. A. Hernandez and S. A. Hertel and G. J. Homenides and M. Horn and D. Q. Huang and D. Hunt and R. S. James and E. Jacquet and K. Jenkins and A. C. Kaboth and A. C. Kamaha and M. K. Kannichankandy and D. Khaitan and A. Khazov and J. Kim and Y. D. Kim and D. Kodroff and E. V. Korolkova and H. Kraus and S. Kravitz and L. Kreczko and V. A. Kudryavtsev and C. Lawes and D. S. Leonard and K. T. Lesko and C. Levy and J. Lin and A. Lindote and W. H. Lippincott and J. Long and M. I. Lopes and W. Lorenzon and C. Lu and S. Luitz and V. Mahajan and P. A. Majewski and A. Manalaysay and R. L. Mannino and R. J. Matheson and C. Maupin and M. E. McCarthy and D. N. McKinsey and J. McLaughlin and J. B. Mclaughlin and R. McMonigle and B. Mitra and E. Mizrachi and M. E. Monzani and K. Morå and E. Morrison and B. J. Mount and M. Murdy and A. St. J. Murphy and H. N. Nelson and F. Neves and A. Nguyen and C. L. O'Brien and F. H. O'Shea and I. Olcina and K. C. Oliver-Mallory and J. Orpwood and K. Y. Oyulmaz and K. J. Palladino and N. J. Pannifer and N. Parveen and S. J. Patton and B. Penning and G. Pereira and E. Perry and T. Pershing and A. Piepke and S. S. Poudel and Y. Qie and J. Reichenbacher and C. A. Rhyne and G. R. C. Rischbieter and E. Ritchey and H. S. Riyat and R. Rosero and N. J. Rowe and T. Rushton and D. Rynders and S. Saltão and D. Santone and I. Sargeant and A. B. M. R. Sazzad and R. W. Schnee and G. Sehr and B. Shafer and S. Shaw and W. Sherman and K. Shi and T. Shutt and C. Silva and G. Sinev and J. Siniscalco and A. M. Slivar and R. Smith and V. N. Solovov and P. Sorensen and J. Soria and T. Stenhouse and T. J. Sumner and A. Swain and M. Szydagis and D. R. Tiedt and M. Timalsina and D. R. Tovey and J. Tranter and M. Trask and K. Trengove and M. Tripathi and A. Usón and A. C. Vaitkus and O. Valentino and V. Velan and A. Wang and J. J. Wang and Y. Wang and L. Weeldreyer and T. J. Whitis and K. Wild and M. Williams and J. Winnicki and L. Wolf and F. L. H. Wolfs and S. Woodford and D. Woodward and C. J. Wright and Q. Xia and J. Xu and Y. Xu and M. Yeh and D. Yeum and J. Young and W. Zha and H. Zhang and T. Zhang and Y. Zhou},
      year={2026},
      eprint={2602.16799},
      archivePrefix={arXiv},
      primaryClass={hep-ex},
      url={https://arxiv.org/abs/2602.16799}, 
}

@misc{liu2022quasiparticlepoisoningsuperconductingqubits,
      title={Quasiparticle Poisoning of Superconducting Qubits from Resonant Absorption of Pair-breaking Photons}, 
      author={Chuan-Hong Liu and David C. Harrison and Shravan Patel and Christopher D. Wilen and Owen Rafferty and Abigail Shearrow and Andrew Ballard and Vito Iaia and Jaseung Ku and Britton L. T. Plourde and Robert McDermott},
      year={2022},
      eprint={2203.06577},
      archivePrefix={arXiv},
      primaryClass={quant-ph},
      url={https://arxiv.org/abs/2203.06577}, 
}

@ARTICLE{2024ApJ...972...70R,
       author = {{Roche}, Cian and {Necib}, Lina and {Lin}, Tongyan and {Ou}, Xiaowei and {Nguyen}, Tri},
        title = "{The Escape Velocity Profile of the Milky Way from Gaia DR3}",
      journal = {\apj},
         year = 2024,
        month = sep,
       volume = {972},
       number = {1},
          eid = {70},
        pages = {70},
          doi = {10.3847/1538-4357/ad58d7},
archivePrefix = {arXiv},
       eprint = {2402.00108},
 primaryClass = {astro-ph.GA},
       adsurl = {https://ui.adsabs.harvard.edu/abs/2024ApJ...972...70R}
}

@article{Burke_2002_Luttinger,
  author  = {Burke, P. J.},
  title   = {Luttinger Liquid Theory as a Model of the Gigahertz
             Electrical Properties of Carbon Nanotubes},
  journal = {IEEE Transactions on Nanotechnology},
  volume  = {1},
  number  = {3},
  pages   = {129--144},
  year    = {2002},
  doi     = {10.1109/TNANO.2002.806823}
}

@article{Burke_2003_RF,
  author  = {Burke, P. J.},
  title   = {An RF Circuit Model for Carbon Nanotubes},
  journal = {IEEE Transactions on Nanotechnology},
  volume  = {2},
  number  = {1},
  pages   = {55--58},
  year    = {2003},
  doi     = {10.1109/TNANO.2003.808503}
}

@misc{Aralis_2026_SQUAT,
  author       = {Aralis, Taylor},
  title        = {{SQUATs}: Qubit-Based Sensors for Dark Matter},
  year         = {2026},
  month        = may,
  howpublished = {Presentation at Superconducting Technologies for Dark
                  Matter, Rome, Italy},
  note         = {Preliminary results}
}

\appendix
\section{End Matter: More details on 1D Luttinger liquid theory}

The purpose of bosonization is to analytically diagonalize an 1D interacting fermion model. Of course, there is no generic cure to the diagonalization of interacting fermion model due to the presence of the quartic interaction term $\sim V_{ijkl}c_i^\dag c_j^\dag c_kc_l$, even in the special 1D scenario. However, in many interesting condensed matter problems, the interaction between fermions (mostly electrons) are sufficiently screened such that regardless of the original form of the interaction, it becomes effectively short-range and can be approximated by some on-site density-density interaction: $\sim \delta(x-x')n(x)n(x')$ (this is why Hubbard models are so widely used in analyzing strongly-correlated condensed matter systems). 

So here comes the heuristic idea of bosonization: the particle density $n(x)$, being bilinear in the fermionic fields, has Bose statistics. That is to say, one might be able to map $n(x)$ to some bosonic field, such that the density-density interaction term becomes quadratic in the bosonic field and is thus easy to diagonalize. 

However, having the interaction terms being quadratic is clearly not enough for the entire Hamiltonian to be diagonalizable analytically – we need the non-interacting part of the Hamiltonian to be still quadratic in terms of the boson fields. Generically, the free fermion Hamiltonian takes the form $\sum_{k,\alpha}\varepsilon_\alpha(k)c_{k,\alpha}^\dag c_{k,\alpha}$, where $k$ is the quasi-momentum and $\alpha$ is the band index. The function $\varepsilon_\alpha(k)$ gives the dispersion relation (band structure) of the band $\alpha$. If we are provided with some arbitrary $\varepsilon_\alpha(k)$, mapping spatial density $n(x)$ to the bosonic fields would lead to some non-quadratic terms (e.g. a $\sim k^2$ term in $\varepsilon_\alpha(k)$ will lead to a cubic term in the bosonic field). Fortunately, in most condensed matter problems we only care about the low-energy effective theory, meaning that we will not be far from the ground state (or Gibbs state under some low enough temperature) – only a vicinity of the Fermi point(s) $E_F = \varepsilon_\alpha(k_F)$ ($E_F$ is the Fermi energy and $k_F$ is the Fermi momentum) will take effect. 

So very commonly $\varepsilon_\alpha(k)$ at around $k_F$ can be approximated by a linear function $\varepsilon_\alpha(k)\simeq v_F k$, where the slope $v_F(\neq 0)$ is called Fermi velocity. This is the case for $E_F$ to be within the range of a band $\alpha$ (the corresponding non-interacting fermion system is a conductor). On the contrary, if $E_F$ is in the gap between two bands (the corresponding non-interacting fermion system being an insulator), or the slope $v_F=0$, we cannot do the above linear approximation. The linear dispersion $\varepsilon_\alpha(k)\simeq v_F k$ is the key for bosonization to take effect: the mapping from density $n(x)$ to the bosonic field keeps the non-interacting part of the Hamiltonian being quadratic in terms of the bosonic field.

With the above road map in mind, let's disclose some details of this magic mapping of bosonization. As mentioned in the main text, a free fermion system in 1D, after zooming into the vicinity of Fermi points $\pm k_F$, involves a right-moving mode $c_R(x)$ with fermi velocity $v_F$ and a left-moving mode $c_L(x)$ with $-v_F$, giving rise to the non-interacting Hamiltonian density
\begin{equation}
    \mathcal{H}_0(x) = -iv_F c_R^\dag\partial_xc_R + iv_Fc_L^\dag\partial_xc_L.
\end{equation}
On top of that, we consider two kinds of on-site density-density interactions: the forward scatterings $\sim n_{R,L}^2(x)$ and the backward scattering $\sim n_R(x)n_L(x)$, described by 
\begin{equation}
    \mathcal{H}_{\text{int}}(x) = \frac{g_4}{2}\left[n_R^2(x)+n_L^2(x)\right] + g_2n_R(x)n_L(x).
\end{equation}
The total Hamiltonian density of the 1D interacting fermion system we have in mind is $\mathcal{H}(x) = \mathcal{H}_0(x) + \mathcal{H}_{\text{int}}(x)$. 

The bosonization map is defined by
\begin{equation}
\begin{aligned}
    &c_R(x) = \gamma_Re^{i\phi_R(x)},\ c_R^\dag(x) = \gamma_Re^{-i\phi_R(x)},\\
    &c_L(x) = \gamma_Le^{-i\phi_L(x)},\ c_L^\dag(x) = \gamma_Le^{i\phi_L(x)}
\end{aligned}
\end{equation}
where $\phi_{R,L}(x)$ are bosonic fields satisfying commutation rules $[\phi_a(x),\phi_b(x')] = \eta_{ab}i\pi\sgn(x-x'),\quad a,b=R,L$ with $\eta_{RR}=1,\ \eta_{LL}=-1,\ \eta_{RL} = \eta_{LR} = 0$, and the Klein factors $\gamma_{R,L}$ satisfying $\{\gamma_a,\gamma_b\} = 2\delta_{ab},\ \gamma_a^\dag = \gamma_a$ guarantees the correct fermionic commutation rules between $c_{R,L}(x)$. Taking the derivative of the commutation rules of $\phi_{R,L}$, one find
\begin{equation}
    [\partial_x\phi_a(x),\phi_a(x')] = \eta_{aa}2\pi i\delta(x-x').
\end{equation}
Recalling that the canonical momentum $\Pi_a$ conjugate to $\phi_a$ is defined by $[\phi_a(x'),\Pi_a(x)] = i\delta(x-x')$, we identify
\begin{equation}
    \Pi_a(x) = -\eta_{aa}\frac{\partial_x\phi_a(x)}{2\pi}.
\end{equation}
With the help of the Baker-Campbell-Hausdorff (BCH) formula, one can further verify 
\begin{equation}
    \left[\frac{\partial_x\phi_a(x)}{2\pi},\gamma_ae^{\pm i\phi_a(x')}\right] = \mp\eta_{aa}\delta(x-x')\gamma_ae^{\pm i\phi_a(x')},
\end{equation}
resembling the commutation relations
\begin{equation}
\begin{aligned}
    &[n_a(x),c_a(x')] = -\delta(x-x')c_a(x'),\\
    &[n_a(x),c_a^\dag(x')] = \delta(x-x')c_a^\dag(x'),
\end{aligned}
\end{equation}
where $n_a(x) = c_a^\dag(x)c_a(x)$ is the particle density in mode $a$. Therefore, we can identify
\begin{equation}
    n_a(x) = \frac{\partial_x\phi_a(x)}{2\pi} = -\eta_{aa}\Pi_a(x).
\end{equation}
A last important result from the bosonization map is 
\begin{equation}
    -ic^\dag_a(x)\partial_xc_a(x) = \eta_{aa}\frac{(\partial_x\phi_a(x))^2}{4\pi},
\end{equation}
which, for example, can be heuristically understood by $\partial_xc_R = \gamma_R\partial_xe^{i\phi_R}\sim ic_R\partial_x\phi_R$ and thus $-ic^\dag_R(x)\partial_xc_R(x) \sim c_R^\dag(x)c_R(x)\partial_x\phi_R(x)\sim n_R(x)\partial_x\phi_R(x)\sim (\partial_x\phi_R(x))^2$. Although this does not give the correct coefficient, we note that all the above relations between fermionic and bosonic languages are exact, which can be rigorously derived from operator product expansion (OPE) in the spatial direction. 

Now, we see that every term in $\mathcal{H}(x) = \mathcal{H}_0(x) + \mathcal{H}_{\text{int}}(x)$ is quadratic in the bosonic fields $\phi_{R,L}(x)$, namely,
\begin{equation}
    \mathcal{H} = \frac{1}{4\pi}\sum_{ab = L,R}V_{ab}(\partial_x\phi_a)(\partial_x\phi_b),
\end{equation}
where $V_{LL}=V_{RR} = v_F + \frac{g_4}{2\pi}$, $V_{RL} = V_{LR} = \frac{g_2}{2\pi}$. By defining $\theta = \frac{\phi_R+\phi_L}{2},\ \varphi = \frac{\phi_R-\phi_L}{2}$, we get the diagonal form
\begin{equation}
    \mathcal{H} = \frac{v}{2\pi}\left[\frac{1}{g}(\partial_x\theta)^2+g(\partial_x\varphi)^2\right],
\end{equation}
where we have defined the Luttinger parameter $g = \sqrt{\frac{2\pi v_F+g_4-g_2}{2\pi v_F+g_4+g_2}}$ and the velocity $v=\sqrt{\left(v_F+\frac{g_4}{2\pi}\right)^2-\left(\frac{g_2}{2\pi}\right)^2}$, and $g<1$ ($g>1$) indicates the repulsive (attractive) interaction between left and right moving modes. The $\theta,\ \varphi$ fields are bosonic fields satisfying commutation rules
\begin{equation}
\begin{aligned}
    &[\theta(x),\varphi(x')] = \frac{i\pi}{2}\sgn(x-x'),\\ 
    &[\theta(x),\theta(x')] = [\varphi(x),\varphi(x')]=0,
\end{aligned}
\end{equation}
such that 
\begin{equation}
    [\theta(x'),\partial_{x}\varphi(x)] = -i\pi\delta(x-x').
\end{equation}
Recalling that the canonical momentum $\Pi_\theta$ conjugate to $\theta$ is defined by $[\theta(x'),\Pi_\theta(x)] = i\delta(x-x')$, we identify
\begin{equation}
    \Pi_\theta(x) = -\frac{\partial_x\varphi(x)}{\pi}.
\end{equation}
This amounts to say that 
\begin{equation}
    \mathcal{H} = \frac{v}{2\pi g}(\partial_x \theta)^2 + \frac{\pi v g}{2} \Pi_\theta^2,
\end{equation}
a massless Klein-Gordon Hamiltonian in 1D with some extra coefficients. 

With the help of canonical equation $\partial_t\theta = \frac{\partial\mathcal{H}}{\partial\Pi_\theta}$, we get $\partial_t\theta = \pi vg\Pi_\theta$ and thus the Lagrangian density is
\begin{equation}
    \mathcal{L} = \Pi_\theta(\partial_t\theta) - \mathcal{H} = \frac{(\partial_t\theta)^2}{2\pi vg} - \frac{v}{2\pi g}(\partial_x\theta)^2.
\end{equation}
The standard Klein-Gordon Lagrangian with wave speed $v$ is
\begin{equation}
    \mathcal{L}_{\text{KG}} = \frac{1}{2v^2}(\partial_t\Phi)^2 - \frac{1}{2}(\partial_x\Phi)^2,
\end{equation}
so our $\mathcal{L}$ inherits this standard form if we define
\begin{equation}
    \Phi = \sqrt{\frac{v}{\pi g}}\theta,
\end{equation}
yielding the familiar Klein-Gordon equation
\begin{equation}
    \frac{1}{v^2}\frac{\partial^2\Phi}{\partial t^2} - \frac{\partial^2\Phi}{\partial x^2} = 0,
\end{equation}
whose solution is
\begin{equation}
    \Phi(t,x) = \int_{-\infty}^{\infty} \frac{dk}{\sqrt{4\pi |k|/v}} \left( a_k e^{-i(\omega_k t - kx)} + a_k^\dagger e^{i(\omega_k t - kx)} \right),
\end{equation}
where $\omega_k = v|k|$. We have adopted the $1/\sqrt{4\pi|k|/v}$ factor such that the canonical commutation relation $[a_k,a_{k'}^\dag] = \delta(k-k')$ holds. 

The above solution gives rise to the two-point function (Wightman propagator) taken with respect to the ground state $\ket{0}$:
\begin{equation}
    \braket{\theta(x,t)\theta(0,0)} = \frac{\pi g}{v}\braket{\Phi(x,t)\Phi(0,0)},
\end{equation}
where
\begin{equation}
\begin{aligned}
    &\braket{\Phi(x,t)\Phi(0,0)} = \iint\frac{dkdk'}{4\pi\sqrt{|kk'|}/v}\\
    &\left(e^{-i(\omega_kt-kx)}\braket{a_ka_{k'} + a_ka_{k'}^\dag}+ e^{i(\omega_kt-kx)}\braket{a_k^\dag a_{k'} + a_k^\dag a_{k'}^\dag}\right).
\end{aligned}
\end{equation}
The only non-zero quantum expectation is the $\braket{a_ka_{k'}^\dag} = \braket{a_{k'}^\dag a_k + \delta(k-k')} = \delta(k-k')$, so
\begin{equation}
\begin{aligned}
    \braket{\Phi(x,t)\Phi(0,0)} &= \int_{-\infty}^\infty\frac{dk}{4\pi|k|/v}e^{-i(\omega_kt-kx)}\\
    &= \frac{v}{4\pi} \left[ \int_{0}^{\infty} \frac{dk}{k} e^{-ik(vt - x)} + \int_{0}^{\infty} \frac{dk}{k} e^{-ik(vt + x)} \right]\\
    &= -\frac{v}{4\pi} \ln \left( \frac{x^2 - (vt - i\epsilon)^2}{\alpha^2} \right),
\end{aligned}
\end{equation}
where we have introduced the small regulator $\eps\rightarrow 0^+$ such that $vt\rightarrow vt-i\eps$, and $\alpha$ is a short-distance cutoff. This gives rise to the two-point function of $\theta$ field:
\begin{equation}
    \braket{ \theta(x,t) \theta(0,0) } = -\frac{g}{4} \ln \left( \frac{x^2 - (vt - i\epsilon)^2}{\alpha^2} \right).
\end{equation}
As a result, the density-density correlation function is
\begin{equation}
\begin{aligned}
    \braket{n(x,t) n(0,0)} &= \frac{1}{\pi^2} \braket{\partial_x \theta(x,t) \partial_y \theta(y,0)}\Big|_{y=0}\\
    &= -\frac{g}{2\pi^2} \frac{x^2 + (vt)^2}{(x^2 - (vt - i\epsilon)^2)^2}\\
    &= -\frac{g}{4\pi^2}\left[\frac{1}{(x-vt+i\epsilon)^2}+\frac{1}{(x+vt-i\epsilon)^2}\right],
\end{aligned}
\end{equation}
which is the same as in \cref{eq: nn_corr}.

\end{document}